\documentclass[preprint]{aastex}
\pdfoutput=1 
\slugcomment{{\sc Accepted to ApJ:} October 3, 2013} 
\usepackage{rotating}
\usepackage{natbib}
\usepackage{lscape}
\usepackage{color}
\tighten

\newcommand{\mjup}{\mbox{$M_{Jup}$}}

\newcommand{\mic}{\mbox{$\mu$m}}
\newcommand{\app}{\mbox{$\sim$}}

\newcommand{\pp}{\mbox{$\pm$}}

\newcommand{\dg}{\mbox{$^\circ$}}

\newcommand{\cand}{\mbox{CD$-$35~2722~B}}

\makeatletter
\newcommand{\Rmnum}[1]{\expandafter\@slowromancap\romannumeral #1@}
\makeatother

\begin{document}

\title {The Gemini NICI Planet-Finding Campaign: \\
The Companion Detection Pipeline.\footnotemark}

\author {Zahed Wahhaj\altaffilmark{2}, 
Michael C. Liu\altaffilmark{3},  
Beth A. Biller\altaffilmark{4},
Eric L. Nielsen\altaffilmark{3},
Laird M. Close\altaffilmark{5},
Thomas L. Hayward\altaffilmark{6},
Markus Hartung\altaffilmark{6},
Mark Chun\altaffilmark{3},
Christ Ftaclas\altaffilmark{3} 
and Douglas W. Toomey\altaffilmark{7}
}

\altaffiltext{2} {European Southern Observatory, Alonso de Cordova 3107,  Vitacura, Casilla 19001, Santiago, Chile}%Me 
\altaffiltext{3} {Institute for Astronomy, University of Hawaii, 2680 Woodlawn Drive, Honolulu, HI 96822}
\altaffiltext{4} {Max-Planck-Institut für Astronomie, Königstuhl 17, D-69117 Heidelberg, Germany}%Beth
\altaffiltext{5} {Steward Observatory, University of Arizona, 933 North Cherry Avenue, Tucson, AZ 85721}
\altaffiltext{6} {Gemini Observatory, Southern Operations Center, c/o AURA, Casilla 603, La Serena, Chile}%hartung
\altaffiltext{7} {Mauna Kea Infrared, LLC, 21 Pookela St., Hilo, HI 96720}%toomey

\begin{abstract}
We present the high-contrast image processing techniques used
by the Gemini NICI Planet-Finding Campaign 
to detect faint companions to bright stars. 
NICI (Near Infrared Coronagraphic Imager) is an adaptive optics instrument installed on the 8-m Gemini South telescope, 
capable of angular and spectral difference imaging and \textcolor{black}{specificially} designed to image exoplanets. 
The Campaign data pipeline achieves median contrasts of 12.6 magnitudes at 0.5$''$ and 14.4 
magnitudes at 1$''$ separation, for a sample of 45 stars 
\textcolor{black}{  ($V=$~4.3--13.9 mag)} from the early phase of the Campaign. 
\textcolor{black}{
We also present a novel approach 
to calculating contrast curves for companion detection based on 95\%
completeness in the recovery of artificial companions injected into
the raw data, while accounting for the false-positive rate.} We use this technique to select the
image processing algorithms that are more successful at recovering
faint simulated point sources. We compare our pipeline to the 
performance of the LOCI algorithm for NICI data and do not find significant improvement with LOCI. 
\end{abstract}

\footnotetext[1]{Based on observations obtained at the Gemini
  Observatory, which is operated by the Association of Universities for
 Research in Astronomy, Inc., under a cooperative agreement with the
  NSF on behalf of the Gemini partnership: the National Science
  Foundation (United States), the Science and Technology Facilities
  Council (United Kingdom), the National Research Council (Canada),
  CONICYT (Chile), the Australian Research Council (Australia),
  Minist\'{e}rio da Ci\^{e}ncia e Tecnologia (Brazil) and Ministerio de
  Ciencia, Tecnolog\'{i}a e Innovaci\'{o}n Productiva (Argentina).}

\section {INTRODUCTION}

The study of exoplanets has advanced admirably in the last two decades,
through the application of radial velocity, transit and microlensing
techniques \citep[e.g.,][]{2008PhST..130a4001M,2010Sci...327..977B,2010ApJ...720.1073G}.
Over \textcolor{black}{800} exoplanet detections have placed significant constraints on the mass and 
separation distribution of exoplanets for separations $\lesssim$3~AU
from \textcolor{black}{their parent} stars \citep[e.g.,][]{2008PASP..120..531C,2010Sci...330..653H}.

In comparison, \textcolor{black}{current} direct imaging techniques can detect Jupiter-mass 
planets at separations beyond $\sim$10~AU from the host star.
Moreover, detecting photons from exoplanets allows us to study their atmospheres
spectroscopically \textcolor{black}{\citep[e.g.,][]{2010ApJ...723..850B, 2011ApJ...733...65B}}. Several extrasolar planets have been
directly imaged, including (1)~HR~8799~bcde, a system with 4 massive
(4--10 \mjup) planets \citep{2008Sci...322.1348M,2010Natur.468.1080M};
(2)~$\beta$~Pictoris~b, a 8$^{+5}_{-2}$ \mjup\ planet first detected
at only 3~AU separation from its primary star
\citep{2009A&A...493L..21L, 2010Sci...329...57L}; (3)~Fomalhaut~b, a
planet with a possible circum-planetary disk detected at 115~AU from
the host star \citep{2008Sci...322.1345K,2013ApJ...775...56K}; (4)~ 1RXS~J1609$-$2105~b,
an 8\pp3\mjup\ planet around a 5~Myr old star in the Upper Scorpius
Association \citep{2008ApJ...689L.153L,2010ApJ...719..497L};
(5) HD~95086~b, a 4--5\mjup\ planet around 10--17~Myr old A8
star \citep{2013ApJ...772L..15R, 2013ApJ...775L..40M};
(6) GJ~504~b, a 4--9\mjup\ planet around a 100-500~Myr old Sun-like
star \citep{2013ApJ...774...11K, 2013arXiv1310.4183J}.    

In recent years, several challenges have been overcome in the field of direct imaging. To achieve high sensitivity to point sources
at sub-arcsecond separations, the intensity of the stellar halo can
now be significantly reduced using adaptive optics (AO) and a coronagraph.
\textcolor{black}{However, long-lived speckles still 
dominate at angular separations close to the star in AO images and
are difficult to distinguish from astronomical point sources.} 
In addition, an impressive variety of instrumental and post-processing techniques have 
been developed to reduce the effect of these speckles. 
For example, angular differential imaging \citep[ADI;][]{2004Sci...305.1442L, 2005JRASC..99..130M} 
decouples the sky rotation of the planet from the speckles. 
Spectral differencing imaging \citep[SDI;][]{1999PASP..111..587R} and spectral deconvolution 
\citep{2002ApJ...578..543S,2007MNRAS.378.1229T} help distinguish between the star and 
planet, taking advantage of the different ways their fluxes change \textcolor{black}{as a function of} wavelength. 
Finally, the study of speckle statistics over time can help to
distinguish planets from speckles over multiple exposures \citep{2009ApJ...698...28G}. 

Several direct imaging AO surveys of young ($\lesssim$100 Myr old)
stars have been completed \textcolor{black}{over the past} decade. These have placed increasingly stronger constraints on the population of
substellar companions and giant planets around nearby stars. 
Null results from these large surveys demonstrate that giant planets
are rare at large \textcolor{black}{semi-major axes} \citep[$\geq$60 AU; ][]{2010ApJ...717..878N}. 
The SDI survey \citep{2007ApJS..173..143B,2008ApJ...674..466N} targeted 
45 nearby young stars, looking for the 1.6\mic\ methane signature expected in cool ($T_{eff} <$ 1400~K) substellar 
atmospheres \citep{1997ApJ...491..856B,2003A&A...402..701B}. The GDPS survey targeted 85 stars, using 6\% methane
filters in the $H$-band with ADI \citep{2007ApJ...670.1367L}.
\citet{2010ApJ...716.1551L} also completed a 58-star survey with the 3.6-m Advanced
Electro-Optical System telescope in Maui. 

\textcolor{black}{
The Gemini NICI Planet-Finding Campaign \citep{2010SPIE.7736E..53L}  is the 
most sensitive large exoplanet imaging survey to date.
At 1$''$ separations, we have achieved a median $H$-band contrast of
$14.4$~mag (this work), compared to $12$~mag for
\citet{2012AA...544A...9V} and  $12.9$~mag for
\citet{2007ApJ...670.1367L}, for example.}    
From December 2008 to April 2012,  the Campaign at the
Gemini-South 8.1-m Telescope targeted a carefully chosen sample of over 200 young nearby stars. 
The Campaign instrument, NICI (Near-Infrared Coronagraphic Imager), was specifically designed for imaging exoplanets
 \citep{2003IAUS..211..521F,2008spie.7015E..49C}. It is equipped with an
85-element curvature adaptive optics system, a Lyot coronagraph, and a
dual-camera system capable of simultaneous spectral difference imaging around the 1.6~\mic\ methane
feature. NICI can also be used in fixed Cassegrain rotator mode for ADI. 

As part of our Campaign observations, we have already discovered cool brown dwarf ($\sim$30-60~\mjup) companions to 4 nearby young stars. 
PZ Tel B is one of very few young ($\sim$10~ Myr old) brown dwarf companions directly imaged at orbital separations
similar to those of giant planets in our own solar system
\citep{2010ApJ...720L..82B}. We have also discovered \cand , one of the
coolest (L4 spectral type) young ($\sim$~100~Myr) brown dwarf
companions directly imaged to date \textcolor{black}{and a member of
  the AB Dor moving group} \citep{2011ApJ...729..139W}. We have found a substellar companion 
in a hierachical triple system, HD~1160 ABC. HD~1160 A is a young
($\sim$ 50~Myr old) A~star and C is a low-mass M3.5~star, and B is the
brown dwarf companion \citep{2012ApJ...750...53N}. 
\textcolor{black}{Lastly, we have resolved the previously known young
  brown dwarf companion to HIP 79797 into a tight (3~AU) binary,
  composed of roughly equal mass brown dwarfs (\app\ 60~\mjup), making
  this system one of the rare substellar binaries in orbit around a
  star \citep{2013ApJ...776....4N}.}
Constraints on the giant planet populations around A~stars, young moving group 
stars and debris disk stars have also been established \citep{2013ApJ...776....4N,2013arXiv1309.1462B,2013ApJ...773..179W}.   

In this paper, we (1)~describe the NICI Campaign data pipeline that was used to achieve 
unprecedented contrasts at sub-arcsecond separations from our target stars
\textcolor{black}{(13.5~mag at 0.5$''$; this work)}; (2)~present 
contrast curves for a set of 45 stars from the early part of the Campaign;
(3)~present a novel technique to compare the results from different pipelines; (4)~compare
the NICI Campaign pipeline to alternate reduction methods; and (5) estimate the astrometric and photometric
precision of NICI Campaign observations.

\section{CAMPAIGN OBSERVING MODES}

\begin{figure}[ht]
  \centerline{
    \hbox {
      \includegraphics[width=18cm]{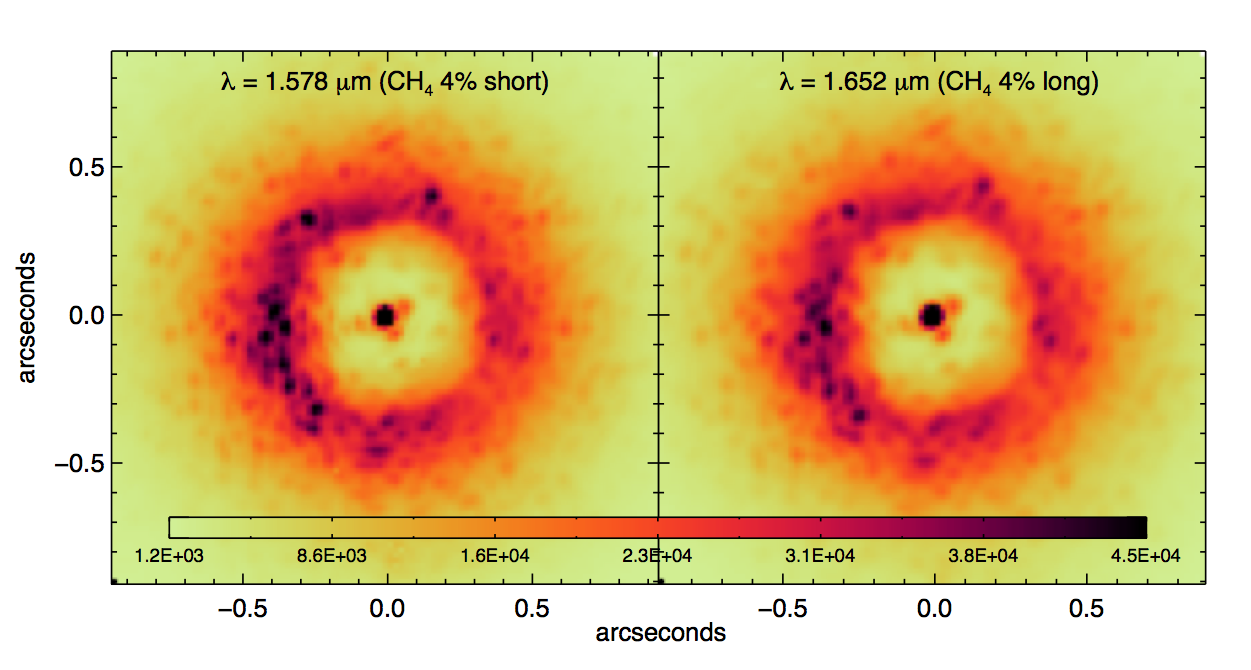}
    }  }
  \caption{NICI Campaign images of the star HD~27290 seen through the focal plane mask, 
simultaneously obtained in the methane filters. The red channel image (right) has been 
rotated to match the sky PA of the blue channel image. The speckle pattern in the red channel is 
very similar to that in the blue channel but is larger in angular scale by roughly 4.6\% (the ratio 
of the central wavelengths of the two filters). The halo in the red channel is also \app5\% fainter, because 
of slightly better halo suppression at that wavelength. The intensity
scale is linear between -9$\times 10^{3}$ and 4.5$\times 10^{4}$ 
counts beyond which the scale ends. The counts in the blue channel image at the stellar peak is 1.1$\times 10^{5}$.}
  \label{fig:ch4_ls}
\end{figure}

Our observing modes have been described in detail in
\citet{2011ApJ...729..139W}. To summarize, we use two modes,
ADI \textcolor{black}{(Angular Difference Imaging)} and ASDI
\textcolor{black}{ (Angular and Spectral Difference Imaging)}, for each target in order to
optimize sensitivity to both methane-bearing and non-methane bearing
companions. In the ASDI mode, we observe simultaneously in two medium-band ($\Delta\lambda/\lambda=$4\%) methane filters, $CH_4$S ($\lambda=$1.578~\mic) 
and $CH_4$L ($\lambda$=1.652~\mic), using NICI's dual camera (Figure~\ref{fig:ch4_ls}). 
In ADI mode, we only image in the $H$-band. 
In both observing modes, the primary star is placed behind a partially
transmissive \textcolor{black}{($\Delta CH_4$S = 6.39~mag)}  
coronagraphic mask with a half-power radius of 0.32$''$. 

%The coronagraphic mask is very effective in suppressing light from the star.
%With adaptive optics, in 95\% (67\%) of the campaign observations, the average intensity of the 
%stellar halo at 0.5$''$  was 358 (716) times fainter
%than the estimated peak intensity of the primary. 

\subsection{The ADI mode}
When using the ADI technique, the telescope rotator is turned off, and the FOV rotates with respect to the instrument detectors. 
This is done so that the instrument and the telescope optics stay aligned with each other and fixed with respect to the detector.
The respective speckle patterns caused by imperfections in the
telescope and instrument optics are thus decoupled from any astronomical companions.
In the de-rotated stack of images, the sky rotations of
individual images are aligned. Thus, the speckle pattern is therefore azimuthally averaged, 
allowing the signal-to-noise of any point source in the field to improve as the square-root of integration time. 
This rate of improvement will only be achieved if the rotational
offset between adjacent images is greater than the FWHM (Full Width at
Half Maximum) of the PSF.
Otherwise, the speckle-dominated regions in neighbouring images in the stack will be correlated and to some extent, the speckles will add as if  they were signal. 
Finally, a reference PSF made by stacking the speckle-aligned images is subtracted from the individual images, so that some fraction of the speckle pattern is removed before it is azimuthally averaged. Thus the noise in the final stacked image is further reduced.   

At large separations from the target star ($\gtrsim$~1.5\arcsec),
our sensitivity is limited by throughput, not by residual speckle structure. 
Thus the benefit of the ADI-only mode is that all the light is sent to
one camera, to achieve maximum sensitivity.
\textcolor{black}{In this observing mode, we choose to acquire} 20 1-minute images using the standard $H$-band filter, which is about four times wider than the 4\% methane filters.

\subsection{The ASDI mode}
To search for close-in planets, we combine NICI's angular
and spectral difference imaging modes into a single unified sequence that we call 
``ASDI''. In this observing mode, a 50-50 beam splitter in NICI
divides the incoming light between the $CH_4$S and $CH_4$L filters which pass the light into the two
imaging cameras, henceforth designated the ``blue'' and ``red'' channels
respectively. The two cameras are read out simultaneously for each exposure and thus the 
corresponding images have nearly identical speckle patterns.
In effect, we have captured the rapidly changing atmospheric component of the speckles. 
However, because of \textcolor{black}{differences in the light path of the two cameras and the wavelength-dependence of the speckles,} 
the red channel PSF is not a perfect match to the blue channel PSF. The long-lived components of the speckle
patterns in both channels are better captured in the remaining images of the full ASDI sequence, from which 
we can construct a reference PSF for further
subtraction. \textcolor{black}{In the ASDI Campaign mode, we typically choose to acquire} 45 1-minute frames.

\subsection{Observing constraints}
Exposure times per frame are kept around 1 minute so that the 
time spent on detector readouts during an observing sequence is small. The hour angles 
when we observe our targets are chosen such that the rotation rate of
the sky is less than 1 degree per minute to avoid excessive smearing
during an individual exposure. The total sky rotation over the observing
sequence is also required to be greater than 15 degrees, to minimize  
self-subtraction of companions at small separations \citep{2009AIPC.1094..425B}.

For most targets, the exposure time per coadd
in the ASDI mode is set small enough so that the star's image behind
the focal plane mask does not saturate. However, if this time is too small  ($\leq$4~secs), 
a striping pattern resulting from electronic interference between the
detectors during readout become prominent in the images. 
Thus for stars with $H<4$ mag, the star behind 
the mask and some part of its halo are allowed to saturate. For these data sets, we obtain
short-exposure (unsaturated) images of the star behind the mask before and after 
the regular sequence. For faint stars, in individual images of 1-minute integrations, the final noise 
contribution from the speckles is smaller than the photon noise in the halo.
So the ASDI mode ceases to yield an advantage. Thus, For stars with  $H>7$ mag, we  
observe only in the ADI mode, but obtain 45 images instead of the usual 20.

\section{DATA REDUCTION}
 
\begin{figure}[ht]
  \centerline{
    \hbox {
      \includegraphics[width=18cm]{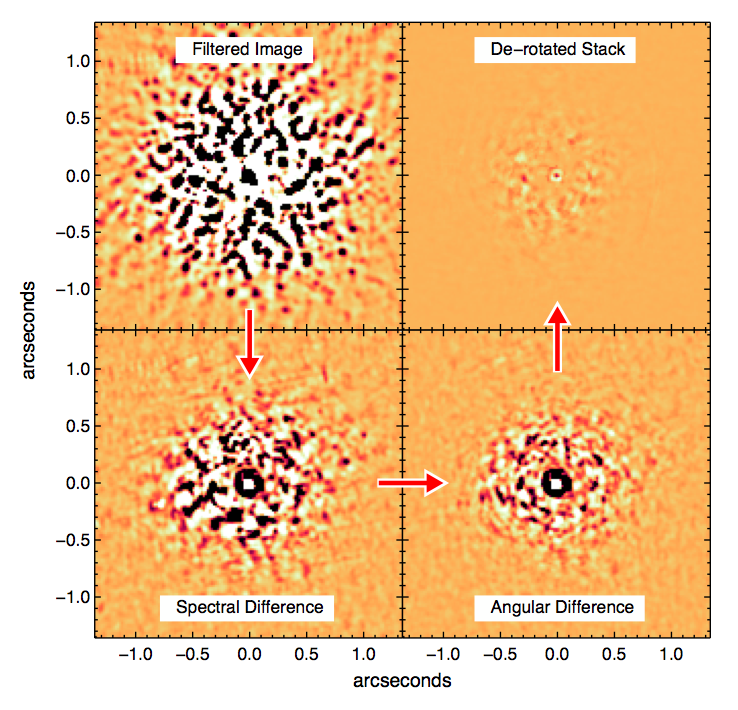}
    }  }
  \caption{Partially reduced images after steps 2 through 6 of the
    NICI pipeline as described in the text.  These images are from the
    reduction of the NICI observation of the star, HD~27290. 
    \textcolor{black}{The faint square pattern, somewhat discernible in the bottom two panels, is due to a striping pattern in the images.} 
    The total integration time for this ASDI observation was 45 minutes.}
  \label{fig:reduc_steps}
\end{figure}

 In our standard ASDI reduction, we reduce the images in six steps (see Figure~\ref{fig:reduc_steps}):

\begin {description}
\item{1.} Do basic reduction: apply flatfield, distortion and image orientation corrections.
\textcolor{black}{
\item{2.} Determine the centroid of the primary star. 
\item{3.} Apply image filters (e.g.\ smoothing).}
\item{4.} Subtract the red channel from the blue channel (SDI).
\item{5.} Subtract the median of all the images from the individual difference images (ADI).
\item{6.} De-rotate each individual image to a common sky orientation and then stack the images. 
\end {description}

\subsubsubsection{Step 1: Basic reduction}
Flatfield images were obtained during most NICI Campaign observing runs, and thus 
most datasets have corresponding flatfields obtained within a few days of their observing date.
The images are divided by the flatfield. We estimate pixel-scale uncertainties of 0.1\%,  
the fractional difference between two halves of a set of flatfield images.  
Bad pixels and cosmic ray hits are replaced by the median of the nearest eight good pixels. The distortion corrections used 
for the NICI data sets are available on the Gemini South NICI website\footnote{http://www.gemini.edu/?q=node/10493}.
The correction is calculated from images of a rectangular
\textcolor{black}{focal-plane} grid mask, upstream of the AO system, which captures the distortion introduced by both the AO system 
and the camera optics. \textcolor{black}{
To estimate the contribution of image distortion to astrometry, we
compared two epochs of Proxima Centauri observations.
There were 73 stars in these NICI images. 
We estimated the astrometric uncertainties to be} 4~mas and 6~mas at separations of 3$''$ and 6$''$ from the center of the detector, respectively (see \S~\ref{section:photast}).
This distortion correction is an improvement over that used in \citet{2011ApJ...729..139W}.

When using ADI, the position angle (PA) of North in the individual NICI images are recorded in the FITS image 
headers at the beginning of the integrations. However, since the instrument rotator is turned off, over the course of 
an individual integration (1 min), the PA changes.  We compute the PA 
value corresponding to the middle of the integration and add it to the FITS headers. This corrected PA is used later when de-rotating 
the individual images to a common orientation.

\subsubsubsection{Steps 2 \& 3: Centroiding and  filtering}
In NICI Campaign datasets, the primary star is usually unsaturated as it is imaged a through a 
coronagraphic mask which is highly attenuating
\citep[$\Delta CH_4$S = 6.39\pp0.03 mag, $\Delta H= $5.94\pp0.05 mag;][]{2011ApJ...729..139W}. 
Thus the location of the primary star in these images is easily determined and then recorded in the FITS headers, to 
be used later for image registration. The centroiding accuracy for unsaturated peaks or peaks 
in the non-linear regime is 0.2~mas (see \S\ref{section:photast}). In ASDI datasets for bright stars (H $<$ 3.5 mag) and 
in most ADI datasets, the primary is saturated. In these images, the peak of the primary is still discernible 
as a negative image. We have estimated that the centroiding accuracy of the saturated images is 9~mas by
comparing these to the centroids of unsaturated short-exposure images obtained right before and after the long exposures.  
\textcolor{black}{For reference, the diffraction-limited FWHM in $H$-band is 43~mas and
the NICI blue-channel plate scale is 17.96~mas.}

Each red and blue channel image in a NICI dataset is processed by a software filter to 
isolate the contribution from point sources (quasi-static speckles and 
real objects) from the more diffuse light of the stellar halo. We describe how we chose the best filter in \S 4. 
In the end, we use the following procedure: (1) convolve the image 
with a Gaussian of 8 pixel FWHM, (2) subtract the convolved image from the 
original image to remove the diffuse PSF light and (3) convolve 
the resulting difference image with a Gaussian of 2 pixel FWHM to suppress the noise on the scale of individual 
pixels. This approach is quite effective at isolating the flux from point sources which have typical FWHM of 3 pixels. 

\subsubsubsection{Step 4: SDI subtraction}
We shift, rotate and \textcolor{black}{demagnify} the red channel images to match the speckles on the corresponding 
blue channel images. The speckles are matched over a 10 pixel wide annulus centered on the primary, 
with inner and outer radii of 0.65$''$ and 0.83$''$, respectively. This region is outside the 
coronagraphic mask, which  is completely transparent at separations greater than 0.55$''$. Five 
parameters are \textcolor{black}{optimized} to find the best match between the two images: (1)~shift in the 
horizontal direction; (2)~shift in the vertical direction; (3)~angular rotation offset; 
(4)~radial demagnification; (5)~intensity scaling. 
The match is determined by minimizing the RMS 
of the pixel values in the annular region in the difference image,  
using a simplex downhill routine \citep[IDL routine AMOEBA; also see][]{1992nrca.book.....P}. We use cubic interpolation 
to estimate the pixel values when rotating and shifting images.
The relative rotation between the two channels is now known to a
precision of 0.1\dg, \textcolor{black}{based on imaging of dense fields}. 
However, to be cautious, we still fit for the best angular offset between the two channels.
\textcolor{black}{For reference, the fractional increase in the RMS of the residual image is 1\%, 5\% and 100\% for angular offset errors of 0.1\dg , 0.2\dg\ and 1\dg , respectively.}

\subsubsubsection{Step 5: ADI subtraction}
Using the pre-computed centroids, all the difference images are registered. 
Then a reference PSF image is created by taking a median combination 
of the registered images. 
\textcolor{black}{ The motion of the star relative to the coronagraphic mask over course of the observation can cause shifts in the speckle pattern. 
Thus, the reference image is matched and subtracted from the individual difference images, optimizing the translational shift and intensity scaling.}
We caution here that a reference PSF subtraction may not only reduce the background RMS, but also remove some of the signal of a real object.
However, such signal loss is significant only when the total sky rotation in an observation sequence results in an 
offset less than about twice the FWHM of the PSF at the angular
separation of the object.     

%We note here that because of the filtering process, the average flux even in the speckle dominated regions is zero 
%in the individual images and thus also in the stacked image. 
%in the same way the red channel images were matched and subtracted from the blue channel images. 
%However, this time, no radial demagnification or angular rotation is
%applied in the fitting procedure. 
%The motion of the star relative to the coronagraphic mask over course of the observation can cause shifts in the speckle pattern. 
%Thus, to be cautious, we fit for best translational shift of the reference image. 

\subsubsubsection{Step 6: De-rotation and stacking}
The double-differenced images (post-SDI and ADI subtraction) are now registered using the pre-computed centroids.
At this step we are interested in aligning astronomical objects, not speckles. 
Also, using the previously computed PAs of the sky orientation, the images are de-rotated to orient North up.
We compute the median of the stack of images to produce the final reduced image. 
At the edges of the field, only a subset of images contribute to the final image, since the square images have 
different sky orientations and thus non-overlapping regions. 
 
\begin{figure}[ht]
  \centerline{
    \hbox {
      \includegraphics[height=9cm]{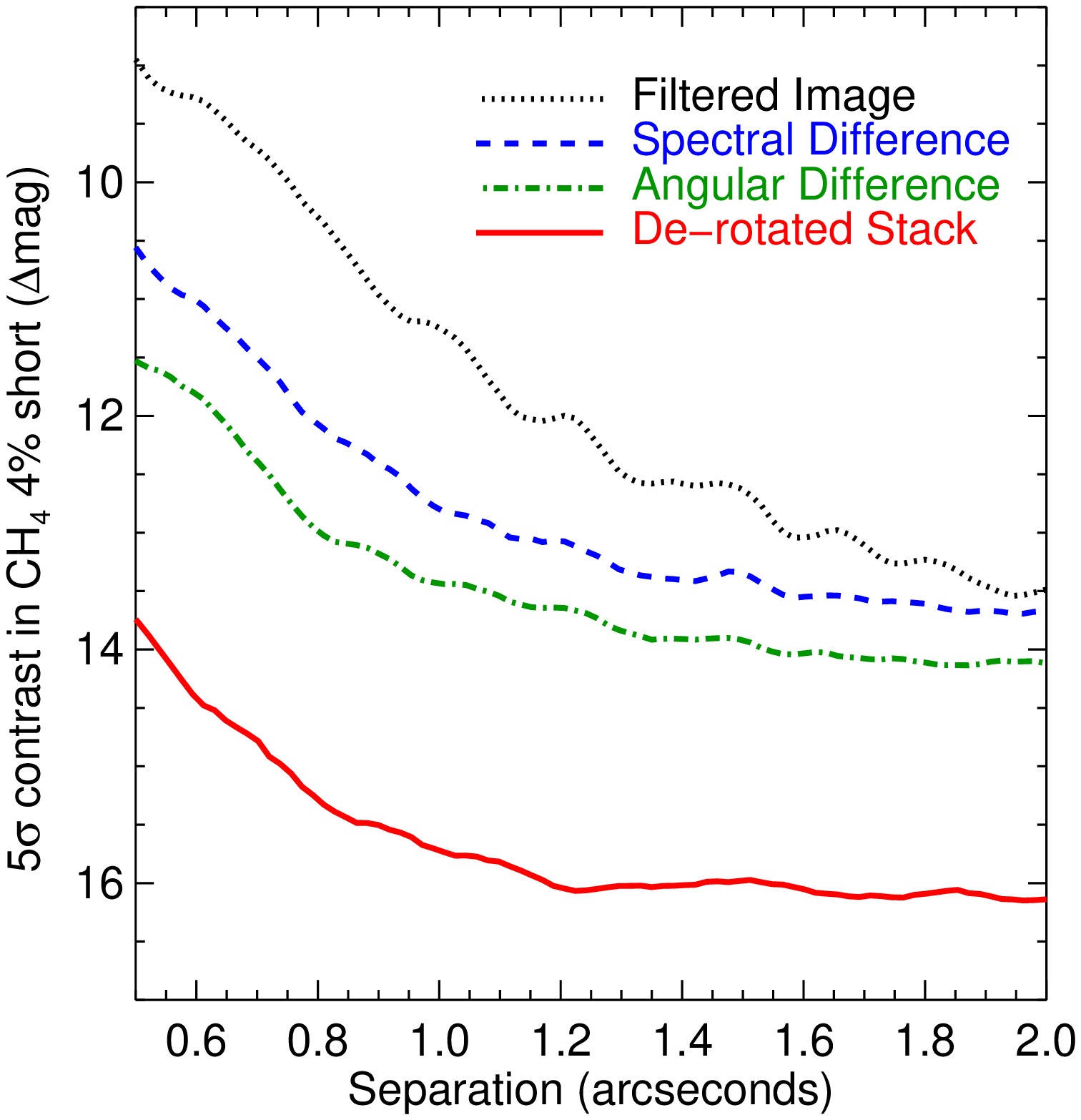}
      \includegraphics[height=9cm]{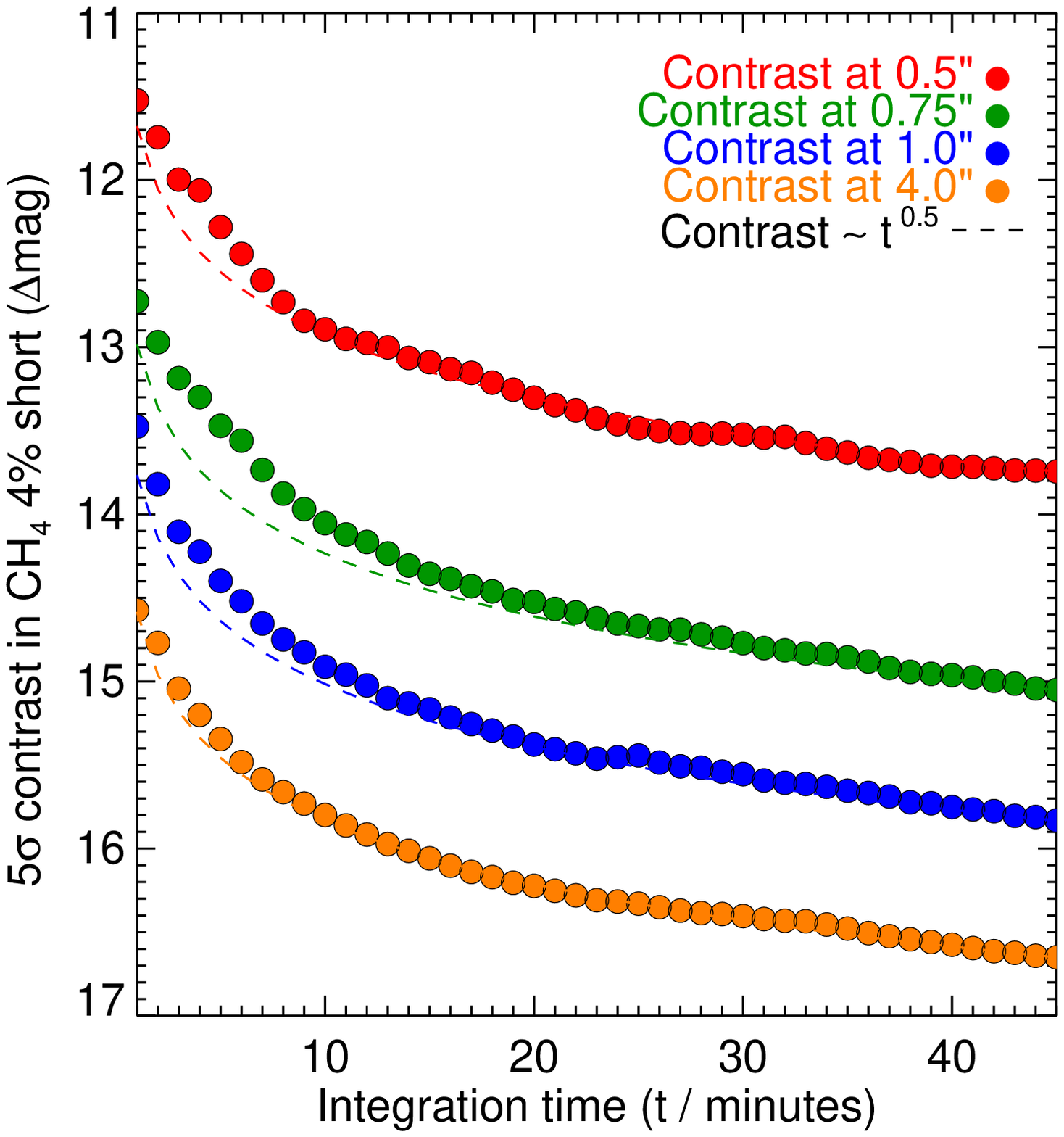}
    }  }
  \caption{Left: Contrast curves (5$\sigma$) after step 2 through 6 of the NICI pipeline as described in the text. The curves computed after steps 3--5 are 
for an individual image, while the curve computed after step 6 represents the contrast limit on the stack of images. 
These contrast curves are from the reduction of the NICI images of the
star HD~27290. The total integration time of this ASDI observation was
45 minutes. Right: Contrast versus integration time at different
projected separations from HD~27290. The contrast improves roughly as $\sqrt{time}$.}
  \label{fig:rcompcon}
\end{figure}

The contrast achieved after each of the reduction steps for the NICI
observation of HD~27290 is shown in the left panel of Figure~\ref{fig:rcompcon}.
\textcolor{black}{ The fact that the improvement in contrast from
  stacking the difference images is roughly equal to the square root
  of the number of images in the stack implies that the residuals
  in the images are not correlated. That there is more than 2$\times$FWHM of
  rotational offset between frames at relevant angular separations is all that
  is necessary for this rate of contrast gain. As we see in the right
  panel of Figure~\ref{fig:rcompcon}, the contrast improves slightly faster
  than the square root of the integration time. This is probably
  because the Strehl ratio improved over the course of the observation.  
The ADI reduction is the same as the ASDI reduction, except that with
ADI we are only dealing with one channel and so we do not perform the SDI reduction in step~4.}

\section{MEASUREMENT OF ACHIEVED CONTRAST VIA SIMULATED COMPANION RECOVERY}
 
In this section, we describe a method to measure the contrast limit of NICI Campaign 
observations using simulated companions. We use scaled versions of the primary as 
seen through the focal plane mask to create these simulated
companions. We have two goals for such work.
The first goal is to measure the contrast at which 95\% of the companions are recovered 
by different reduction pipelines (see \S~5 \& 6), and thereby identify
the superior pipelines. 
\textcolor{black}{We chose the 95\% limit because, as will be clear later, an estimate of
the 99\% completeness limit  takes roughly 5 times longer to compute. 
The second goal is to estimate the systematic and random errors introduced
into the photometric and astrometric properties of the images by the
reduction process.} 
The method to do this is conceptually described below and then in greater detail in subsequent subsections.

Our method is to reprocess each data set with two \textcolor{black}{modified} reductions.  The first reduction is to
purposefully mis-rotate the frames so that the result does not contain any companions  (a source-free reduction).
This allows us to \textcolor{black}{(1)~set an upper contrast limit above which no
detections (false positives) will be accepted and (2)~determine the nominal 1$\sigma$ 
detection-limit curve.
For the second reduction, we introduce fake companions 20 times brighter
than the previously determined 1$\sigma$ limit (the 20$\sigma$ limit).
The photometric and astrometric properties of the recovered companions
of known contrast from the second reduction 
provide an estimate for the systematic and random errors introduced by the pipeline.}
Finally, the recovered companions are reinserted into the source-free reduction at several 
locations, and scaled up in intensity (starting from zero flux) until they meet our detection criteria. 
\textcolor{black}{The contrast (1$\sigma$ limit times the scale factor)
  at which 95\% of the simulated companions are detected (95\%
  completeness)  establishes our contrast curve for that star. }

%In the second reduction, we insert the 20$\sigma$ fakes into the individual images and then reduce the 
%data again in the same way. 
%The recovered companions allow us to measure the systematic and random errors introduced 
%into their photometry and astrometry by the pipeline.

\subsection{Making a source-free image to determine the false-positive threshold}
To remove any real objects in a dataset from the final reduced image, in step 6 of the reduction procedure, 
we artificially set the orientation of the images to thirty times their actual sky PA values.
Since the the directions of North on the images are now \textcolor{black}{incorrect},
when the stack of images \textcolor{black}{is oriented to a common PA and} median-combined
any real objects disappear. Moreover, we can now \textcolor{black}{assume} that any point-source detection in this reduction is 
a false positive and that valid detections must be at a higher sigma level than the strongest false-positive. 
\textcolor{black}{This is the first criteria that a companion detection
  must satisfy (in the properly oriented and stacked images) to
  be considered real by our algorithm.} 
%Thus we establish a single sigma threshold for the entire image, or equivalently an $N\sigma$ contrast curve
%as the false-positive limit.

\subsection{1$\sigma$ contrast curve measurement} 
Since the focal plane mask attenuation is measured 
using a photometry aperture of 3 pixel radius \citep{2011ApJ...729..139W}, the contrast curve is also measured 
using apertures of the same radius. We create an image from the final source-free reduced image, replacing 
each pixel value by the aperture (radius = 3 pixels) flux centered on that pixel. We then calculate the noise as a 
function of radius in this aperture flux map from the standard deviation 
of all pixel values in an annulus.  Thus, the 1$\sigma$ contrast curve is the product of: (1)~the inverse of the noise curve, 
(2)~the flux of the primary (seen through the mask) and (3) the mask attenuation. 

\subsection{Detection criteria} 
To extract sources from the reduced image, we create a signal-to-noise map. To do this, we divide 
each pixel in the reduced image by the standard deviation of all the pixel values in a narrow annulus containing the chosen pixel. 
Since there are no real sources in the signal-to-noise map of the source-free image,  the counts in 
the brightest spurious detection set the threshold for believable detections. To detect the spurious point sources, 
we first convolve the signal-to-noise map 
by the star spot (a circular region of radius 5 pixels centered on the star).
This is done so that realistically shaped point sources, as opposed to single bright pixels or an  
accidental cluster of bright pixels, are preferentially amplified in the signal-to-noise map.  
As an initial sample of possible detections, we then take all the pixels brighter than three 
times the pixel-to-pixel standard deviation of the
\textcolor{black}{convolved} signal-to-noise image. 
Next, we shift the star spot to the same center as each of the
putative detections and scale its flux to match 
the peak pixel values \textcolor{black}{in the unconvolved map}. We then calculate a detection-strength metric given by $flux/\chi_\nu^2$, where $flux$ is the 
aperture (3 pixel radius) ``flux'' of the putative detection in the \textcolor{black}{unconvolved} S/N map, and $\chi_\nu^2$ is the reduced $\chi^2$ from the 
comparison between the detection and the star spot within the aperture. 
The flux is included in this metric to distinguish strong detections from faint point sources at the 1$\sigma$ level 
which can also have $\chi_\nu^2$ close to unity.
The error in the pixel values is assumed to be one, since this is a signal-to-noise map. 
We find that point sources with a detection-strength of $\geq$0.25 look credible to the eye. 
This is satisfactory because Gaussian PSFs with peak intensity 5 times greater than the 
RMS of a Gaussian background also give a detection strength of 0.25. Finally, for a detection 
to be considered real, it must satisfy both this detection-strength criteria and be above the no-false-positive threshold.

\subsection{Insertion of  simulated 20$\sigma$ companions}

We insert simulated companions at the 20$\sigma$ level to estimate the random and systematic effects of the pipeline and also 
to measure recovery completeness. 
After basic reduction of the individual images (step 1 of the pipeline), we embed 20$\sigma$ simulated 
companions into the individual blue channel images, using our 1$\sigma$ contrast curve as reference. 
The simulated companions are assumed to be methane-rich, such that they have 
negligible flux in the red channel. Note that our final contrast curves will be adjusted later 
to account for the methane channel brightnesses for companions as a
function of spectral types (\S\ref{section:contrast_adjust}). 

The assumed sky PAs for the red and blue channel images are set to the same artificial values as in the source-free 
reduction, so that this final reduced image will be same, except for the embedded companions. 
The 20$\sigma$ companions are placed in each image at separations of 0.36$''$ to 6.3$''$ (20 to 350 pixels), 
with the first companion at 0.36$''$ separation and zero PA. Each subsequent companion is placed farther out 
by 0.09$''$ (5~pixels) and at a PA increased by 50 degrees (Figure~\ref{fig:sim-comp-pattern}). After the standard ASDI or ADI reduction procedure is applied and 
the signal-to-noise map is constructed, we easily recover all the simulated companions as roughly 20$\sigma$ detections.
However, all the companions are recovered at different slightly contrasts than their original values and possibly have 
different fluxes depending upon the local effects of the reduction process. 
Our approach of generating the contrast (described below) naturally
incorporates these effects.

\subsection{Derivation of the 95\% completeness contrast}
\begin{figure}[ht]
  \centerline{
    \hbox {
      \includegraphics[width=18cm]{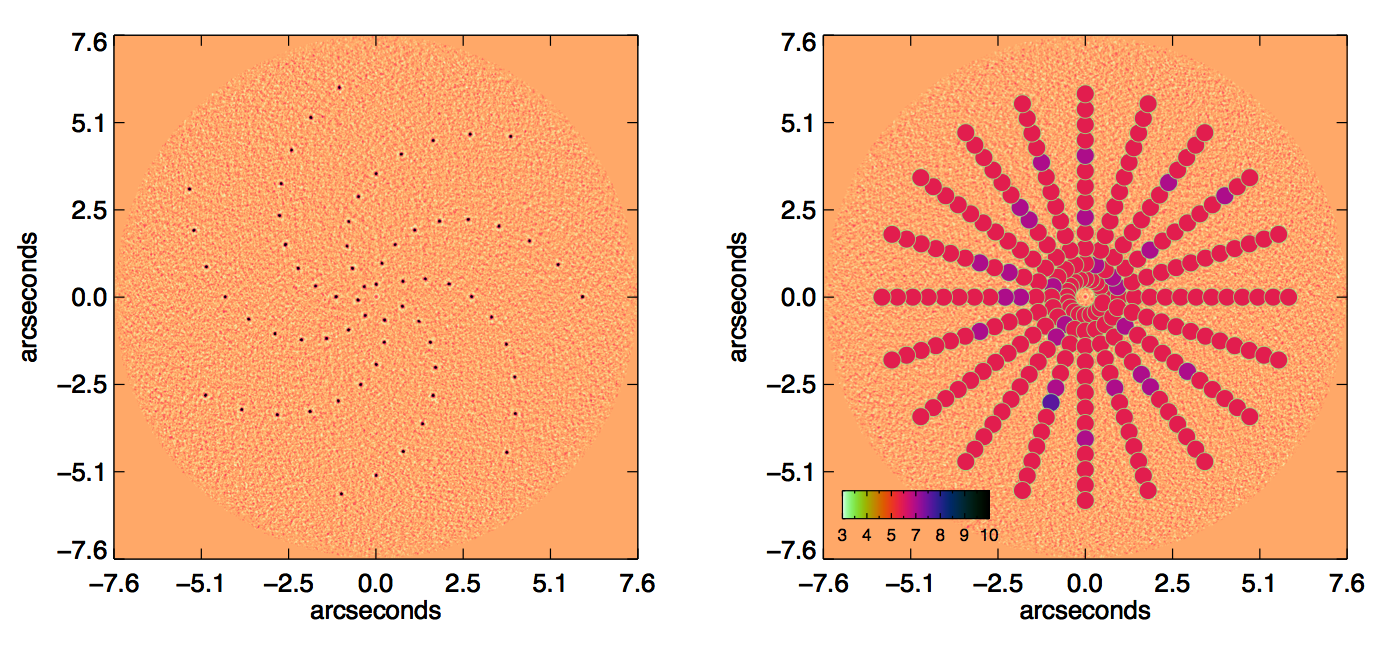}
    }  }
  \caption{Left: Signal-to-noise map showing the recovery of simulated companions 20 times brighter than the 1$\sigma$ contrast curve. Right: 
Every 5$^{th}$ ring of the 67 separation rings and 20 PAs at which the simulated companions were reinserted and recovered are shown with colored circles. The colors indicate the smallest signal-to-noise at which the companions satisfied the detection criteria. The color bar indicates the signal-to-noise scale. The image shown is the signal-to-noise map of the source-free reduction made from UY Pic data set of UT Dec 4, 2009.}
  \label{fig:sim-comp-pattern}
\end{figure}

The 20$\sigma$  companions in the signal-to-noise map at each separation (67 in total) are then cut out 
and inserted into the signal-to-noise map of the source-free reduction at 20 different PAs. 
At each location, we decrease the flux of the companion until it
fails to satisfy our detection criteria.
\textcolor{black}{
Remember that the two criteria are: (1) the source is above the 
false-positive threshold determined from the S/N map of the source-free image
and (2) the source has good detection-strength which depends upon its
total ``flux'' in the S/N map and PSF shape.
Using these test insertions at each location, we record the maximum contrast (smallest companion flux) which resulted in detection. 
Thus, we know the contrast limit for each location. For each
separation, we determine the contrast at which 19 out of 20 (95\%)
companions are recovered.}  
The resulting curve represents the 95\%  completeness contrast. 
The right panel of Figure~\ref{fig:sim-comp-pattern} shows the locations at which the recovered 20$\sigma$ 
companions were reinserted, i.e., where the contrast limits were determined.  The standard contrast curve throws away information
about the azimuthal variation of the contrast achieved. Figure~\ref{fig:sim-comp-pattern} shows how much azimuthal variation there is in a typical 
reduction. The images shown are from the UY Pic ASDI data set obtained on 2009 December 4 UT.  

\section{OPTIMIZING IMAGE FILTERING VIA SIMULATED COMPANION RECOVERY}

\subsection{Description of the image filters}
\label{section:image_filters}
A better reduction should yield a better 95\% completeness curve.
Thus we use the 95\% completeness curves to judge the effectiveness of different image filters 
in step 2 of the standard reduction procedure.
The image filters we applied were different combinations of the following basic processes:
``u'~unsharp-masking; ``c''~'catch' filtering (unsharp masking + smoothing); ``a''~removal of azimuthal 
profile and ``d''~de-striping. Unsharp masking is done 
by convolving an image with a Gaussian of 8-pixel FWHM and subtracting this convolved image from the original. 
Catch filtering entails a further smoothing with a Gaussian of 2-pixel
FWHM in order to suppress pixel-to-pixel noise.
\textcolor{black}{The 2- and 8-pixel spatial scales are half and double
  the size, respectively, of the FWHM of the PSF in the poorer half of our
  data sets. The goal is to isolate to point sources.}
Azimuthal profiles were removed by a procedure described in \citet{2011ApJ...729..139W}. The procedure is very effective
at removing PSF features which are extended azimuthally.   
De-striping is used to remove the striping pattern caused by detector electronics which usually occurs in short exposures.
It is only applied to the region where the flux from the primary is fainter than twice the local pixel-to-pixel RMS. 
In this exterior region of the images, we subtract from each pixel the median of the closest 30 closest 
pixels in the same row, and then in the same column.  

\subsection{Comparison of the image filters}
\begin{figure}[ht]
  \centerline{
    \hbox {
      \includegraphics[height=8cm]{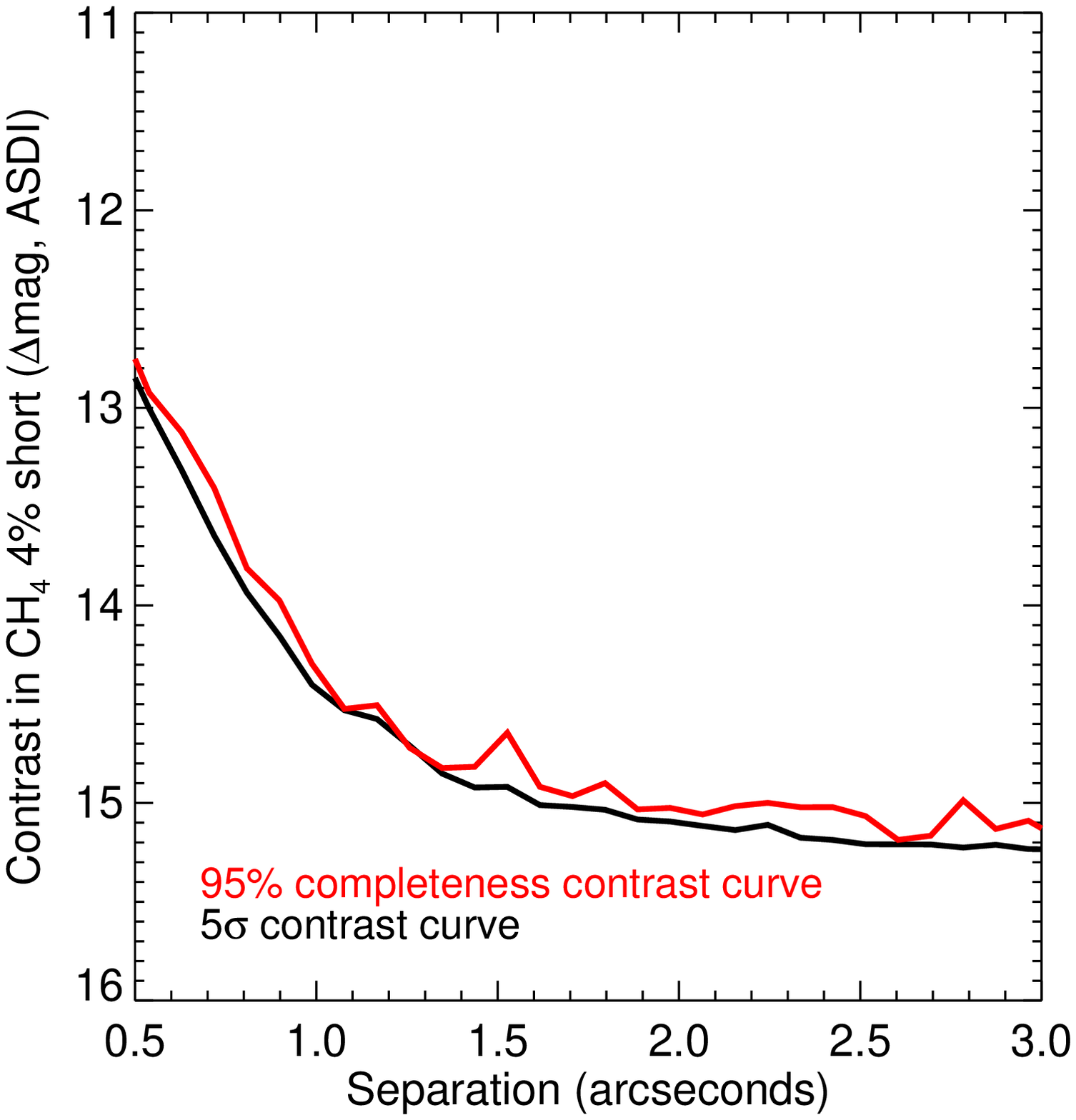}
      \includegraphics[height=8cm]{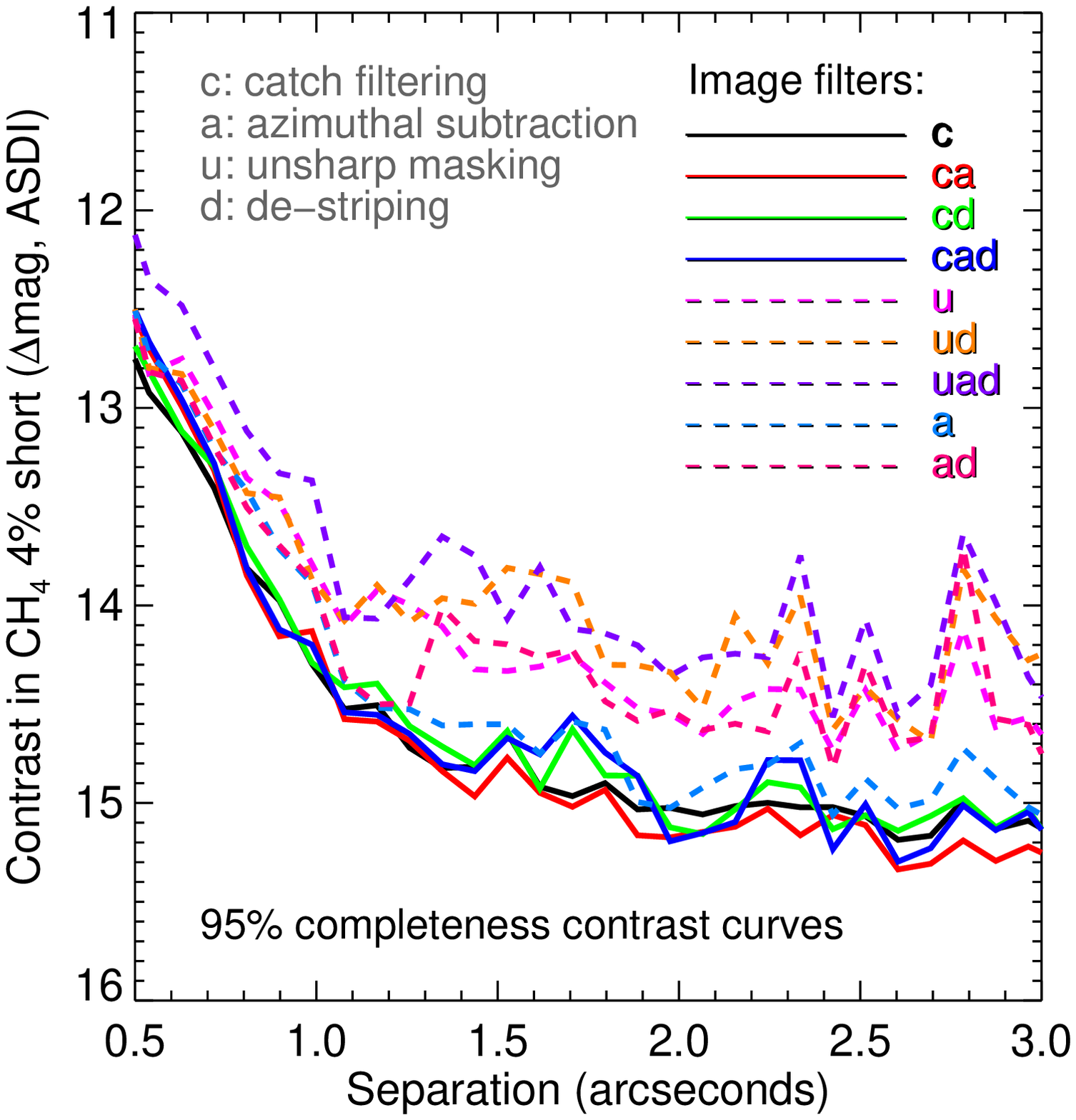}
    } 
  }
  \caption{Left: The 95\% completeness contrast curve (red line) compared to the 5$\sigma$ contrast curve (black line) as computed for the UY Pic 2009 Dec 4 data set. Right: The 95\% completeness contrast curve for the same data set for several different image filters tested in the NICI pipeline. The different image filters are described in \S\ref{section:image_filters}}
\label{fig:concurve}
\end{figure}

Figure~\ref{fig:concurve} shows the 95\% completeness 
contrasts for different combinations of image filters used in the reduction of our UY Pic 2009 dataset. 
The filter 'cd' indicates catch filtering followed by de-striping. We can see that the 'c' filter yields the highest 
contrast close to the star, while at larger separations it yields one
of the highest contrasts and also yields a smoother curve. The roughness of the other curves suggests that the other filters may be
responsible for residual structures which are resulting in a patchy noise floor. 
The filter combinations that involve 
catch filtering yield better contrasts even at large separations where striping or sky noise dominates.
It seems likely that the small-scale smoothing involved in catch filtering mitigates the problem 
of residual structures forming \textcolor{black}{more complex} patterns 
in the de-rotated reduced stack. At smaller separations, catch filtering leads to better fitting 
of the speckle pattern during the subtraction process. As expected, we found that the de-striping 
filter is more useful for datasets where the integration time per coadd is short, because the striping is more severe 
in these cases.  
We also make the following observations: (1) the best filters are 'c', 'ca', 'cd' and 'cad', and they are more or less
equivalent; (2) the 'a' and the 'd' filters can slightly worsen
performance when used together and (3) the catch filter 
generally improves performance. We choose the 'cad' filter for the pipeline, 
as insurance against cases where the striping pattern or azimuthal variation across images is unusually severe.    
In Figure 5, we see that the 95\% completeness curve agrees well with the nominal  
5$\sigma$ noise curve for the UY Pic dataset.
  
\section{COMPARISON WITH THE LOCI ALGORITHM}
Algorithms for processing \textcolor{black}{high contrast imaging} have steadily 
improved at removing starlight while preserving the planet's flux. 
Currently, the LOCI (Locally Optimized Combination of Images)
algorithm is among the most commonly 
used post-processing method \citep{2007ApJ...660..770L}. 
It was originally used to process ADI data from the Gemini Deep Planet Survey \citep{2007ApJ...670.1367L}.
LOCI builds the optimal PSF for any given sub-region of the science
image, \textcolor{black}{choosing linear combinations of} science images where the PSF is very similar. 
This algorithm has undergone several improvements. For example, the subtraction region can be masked so as not to bias
the construction of a reference PSF by real companions \citep{2010SPIE.7736E..52M}. Moreover, the subtraction region can reduced to one pixel, and LOCI 
choices tuned to improve point-source sensitivity
\citep{2011ApJ...741...55S}. Lastly, LOCI \textcolor{black}{can be
  applied} to integral field spectroscopy observations, where the reference PSF can be built
 from images of the star over an entire spectral and temporal range \textcolor{black}{\citep[e.g.,][]{2011ApJ...733...65B}}.  \citet{2012ApJS..199....6P} have added the innovation that LOCI is programmed not just to minimize the local noise in the images but also to preserve the local flux. In other words, the algorithm tries to  maximize the local signal-to-noise. 

We can now compare the effectiveness of the original LOCI algorithm to
that of our standard \textcolor{black}{NICI Campaign} pipeline 
using their respective 95\% completeness contrast curves. The more
sophisticated versions \textcolor{black}{of LOCI} \citep{2011ApJ...741...55S,2012ApJS..199....6P} designed for improved flux 
conservation are quite computationally intensive, and we do not test with them in this paper.  

\begin{figure}[ht]
  \centerline{
      \includegraphics[height=12cm]{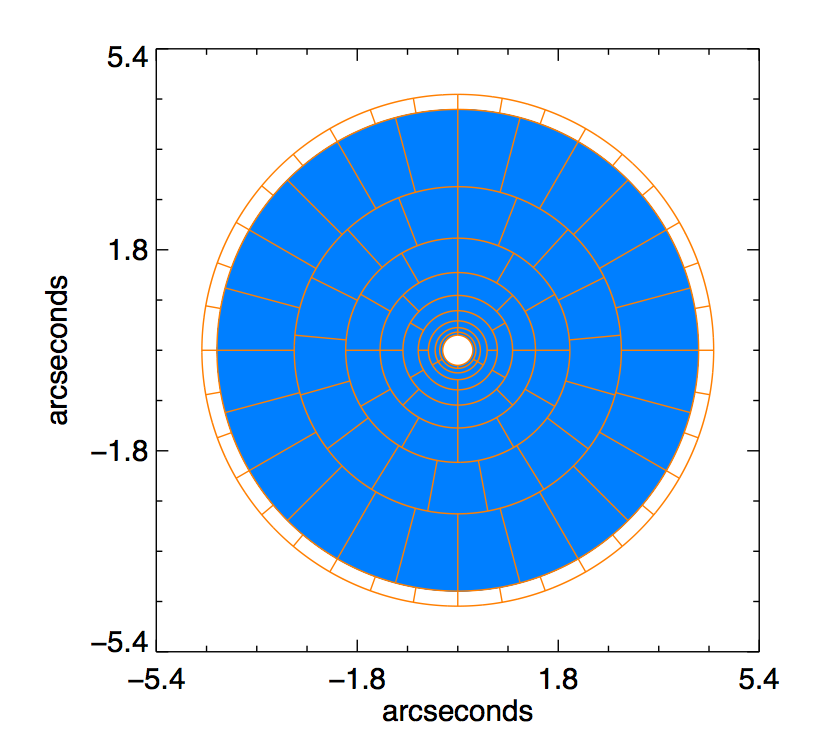}
    }
  \caption{The subtraction and reference sectors for an example set of parameters in the LOCI algorithm, with $dr$ 
increasing at a rate of 50\% ($dr_{fac}$=1.5) and number of sectors linearly increasing with 
radius, from 3 to 18. The reference regions extend beyond the subtraction regions radially 
by 15 pixels ($dw=$15). This extension is shown without color for the outermost ring. }
  
\label{fig:loci_sectors}
\end{figure}

The LOCI algorithm is applied in step 5 of our pipeline, after we subtract the red channel  images from the blue channel images. In short, the algorithm works likes this. When processing each science image, instead of subtracting the median combination of the other images in the dataset, we subtract a linear combination of the images optimized for each region of the image. These regions are configured as segmented concentric annuli around the primary star  (Figure~\ref{fig:loci_sectors}). For each region that we want to optimally subtract speckles from, we define a reference region which extends outward radially from the subtraction region by an extra 15--30 pixels. The linear combination of images is optimized over the larger reference region to 
minimize the RMS of the subtraction. 

The LOCI algorithm \textcolor{black}{can be tuned} by the following 
parameters: $N_\delta$, $N_A$, $dr$, $g$ and $dw$. $N_\delta$ represents the minimum offset 
required when selecting reference frames, measured as arclength traced by the sky rotation 
divided by the PSF FWHM.  $N_A$ is the \textcolor{black}{area} of the reference region in units of the PSF FWHM, 
and $dr$ is  the radial width of the subtraction region in units of PSF FWHM. The parameter, $g$, 
is the ratio of the radial extent to the angular extent of the reference region, and $dw$ is the extra 
radial width of the reference region beyond that of the 
subtraction region.  \textcolor{black}{For our LOCI processing, we 
assume a PSF FWHM of 3 pixels, which is representative of our NICI data.} 

Instead of setting $N_A$ and $g$ directly, we set it through two alternate parameters 
which are easier to conceptualize. We set the number of sectors in the annulus at $r=7''$ using 
by the parameter, $N_{sec}$. We choose the number of sectors per
annulus to decrease linearly with 
decreasing radius, \textcolor{black}{rounding the number to nearest integer}. We increase the radial width of the subtraction regions geometrically 
going outwards by the factor $dr_{fac}$ with each annular increment. The $N_A$ and $g$ values corresponding 
to some of the $N_{sec}$ and $dr_{fac}$ values  we used are given in Table~\ref{tab:loci_prams}. 
Our choice of sectors for an example set of LOCI parameters is shown in Figure~\ref{fig:loci_sectors}. 
 
\begin{figure}[ht]
  \centerline{
    \vbox{
      \hbox{
        \includegraphics[height=8cm]{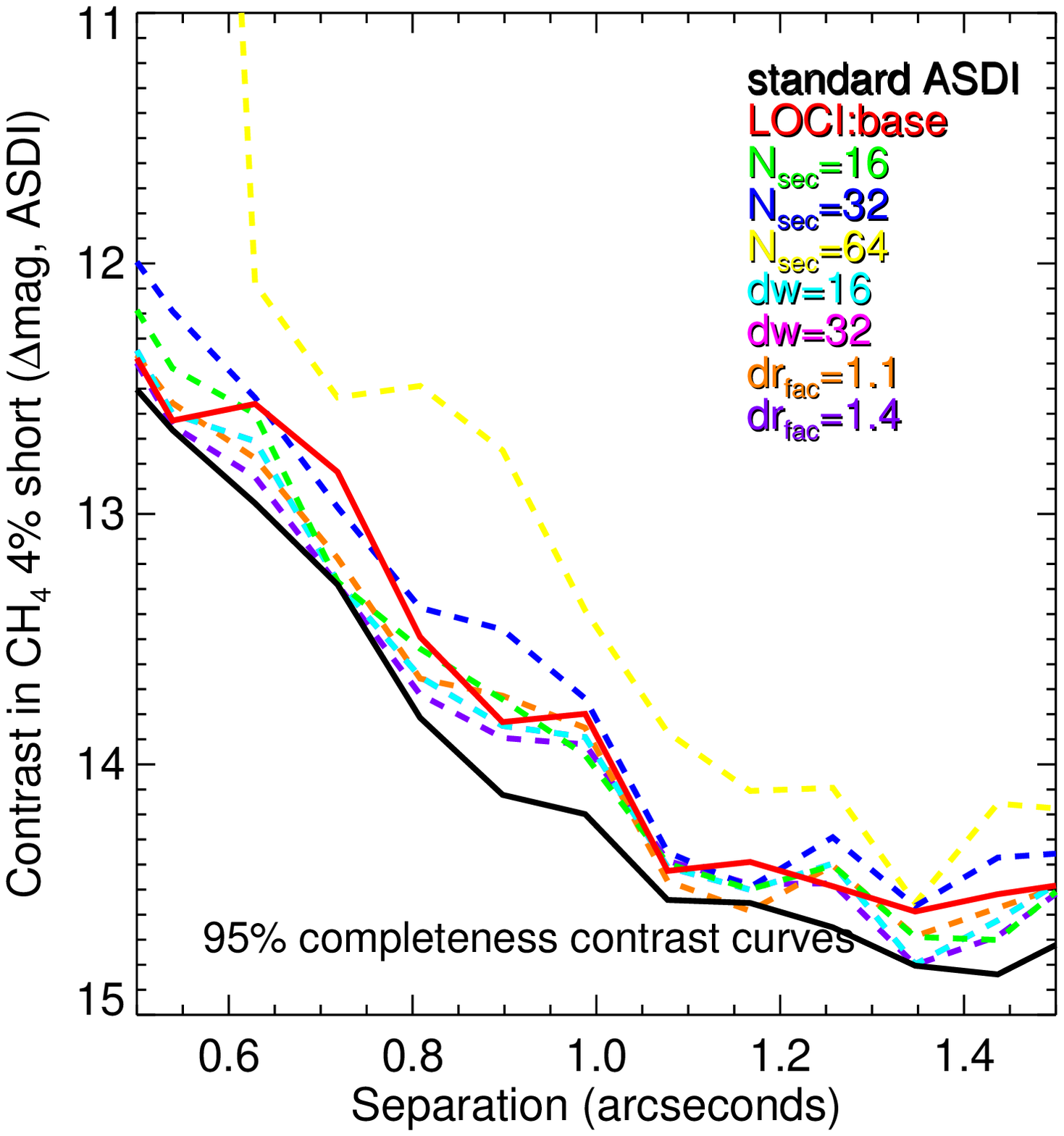}
        \includegraphics[height=8cm]{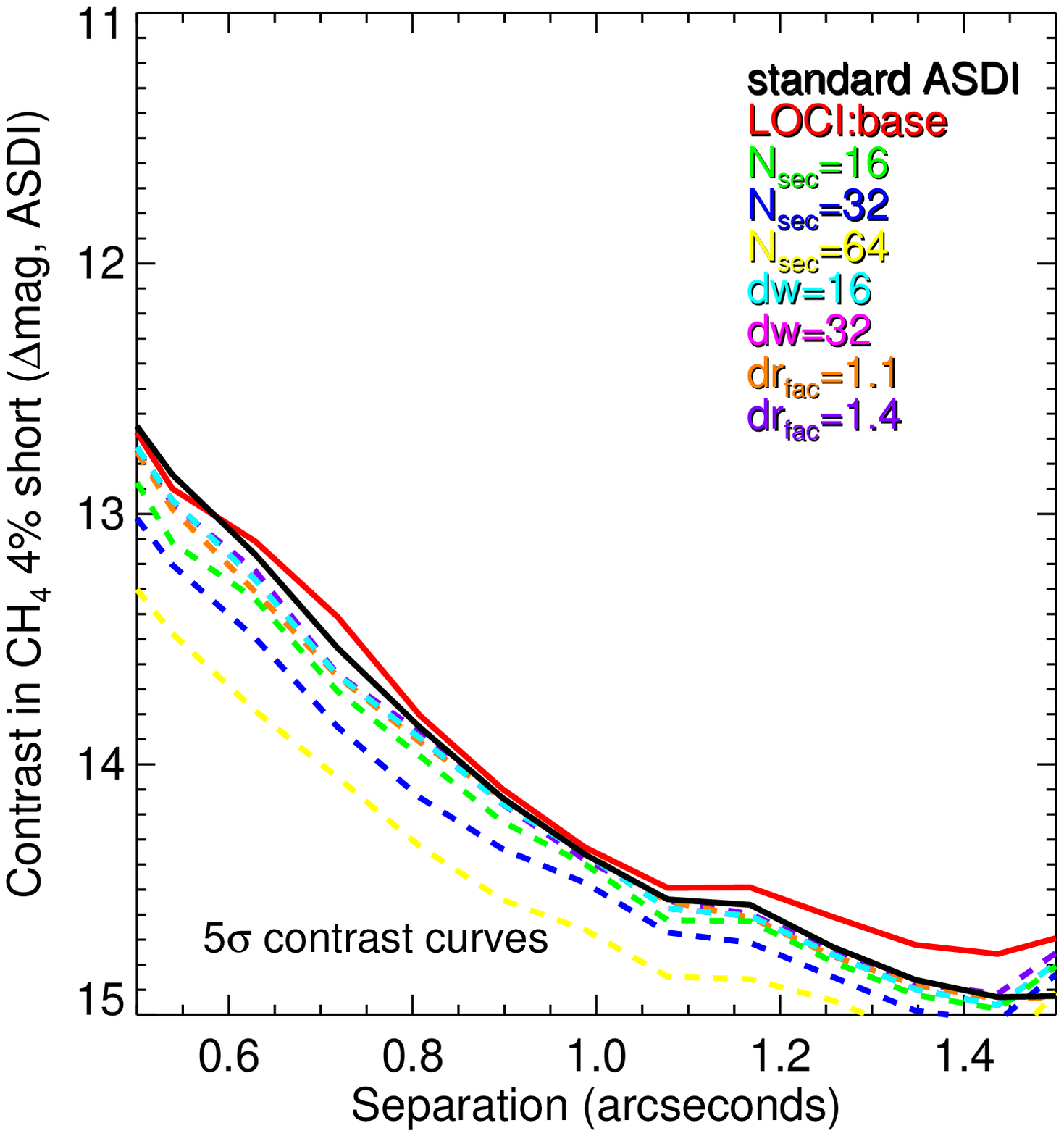}
      }
      \hbox{
        \includegraphics[height=8cm]{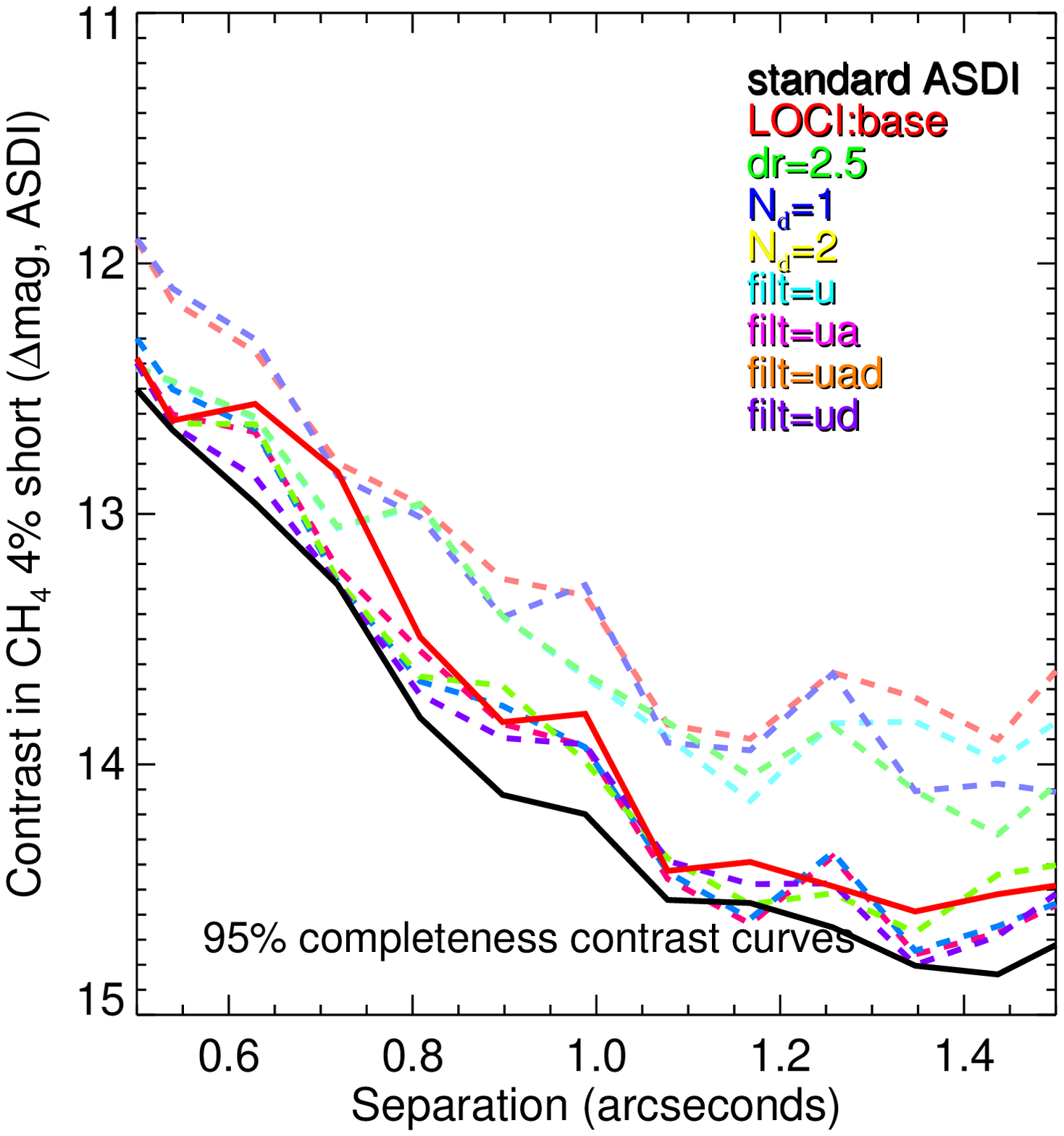}
        \includegraphics[height=8cm]{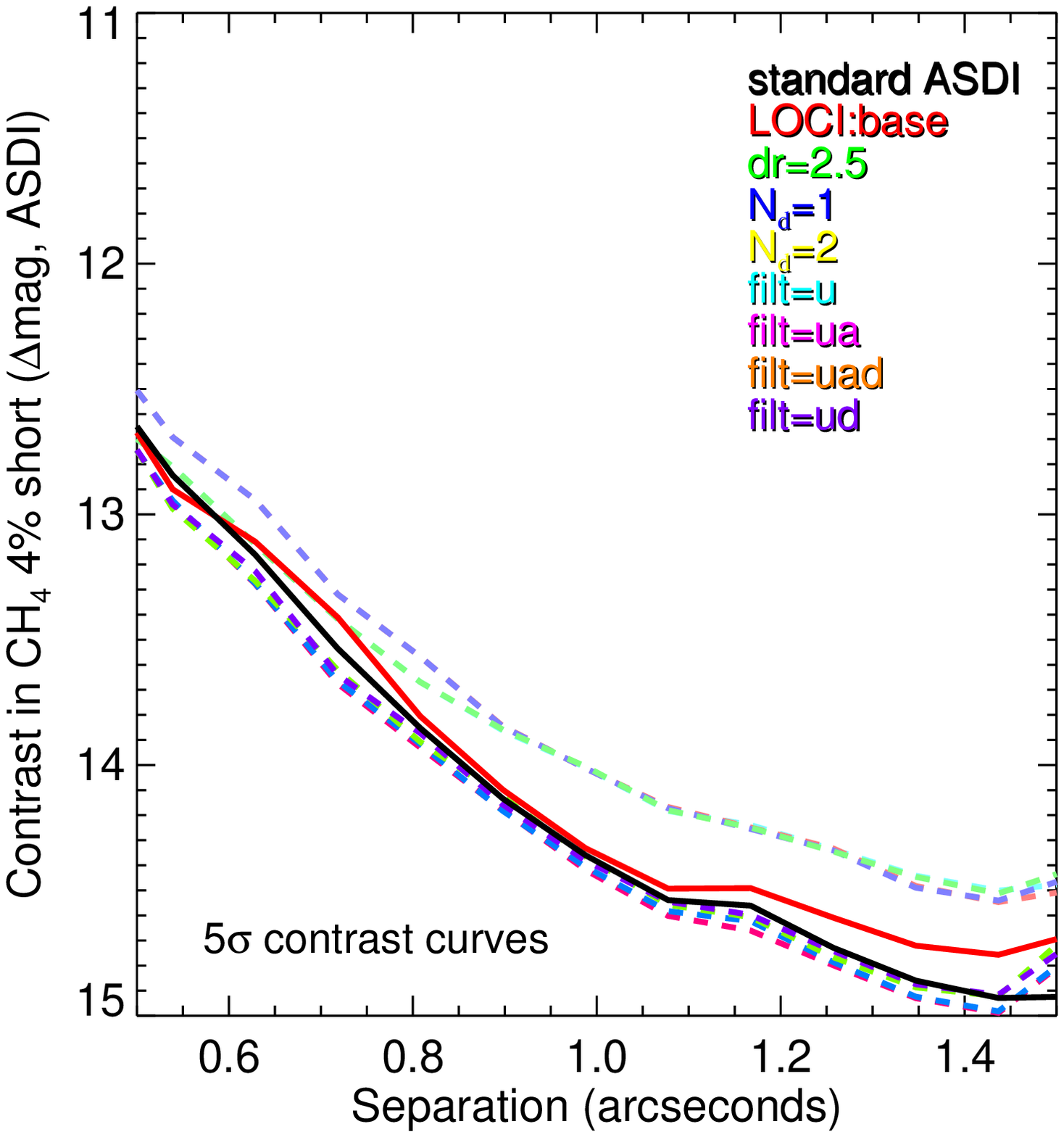}
      }
    }
  }
  \caption{Left (top and bottom): Comparison of 95\% completeness contrasts for different LOCI reductions and 
\textcolor{black}{our standard ASDI reduction for UY Pic two-channel ASDI 2009 data}. The LOCI base parameters were set to 
$N_{sec}=8$, $N_{\delta}=3$,  $dr=4$,  $dr_{fac}=1.2$, $dw=1$ and filt = 'cad' (see text for explanation).  
For the other LOCI reductions, we set only one parameter to 
a different value as noted in the legends. Also shown is the 95\% completeness contrast 
(black line) for the standard ASDI algorithm. The LOCI reductions are 
not an improvement on our standard ASDI reduction. Right (top and bottom): Comparison 
of the nominal 5$\sigma$ contrast curves.}
\label{fig:loci_asdi_95p_uypic}
\end{figure}

\begin{figure}[ht]
  \centerline{
      \includegraphics[height=8cm]{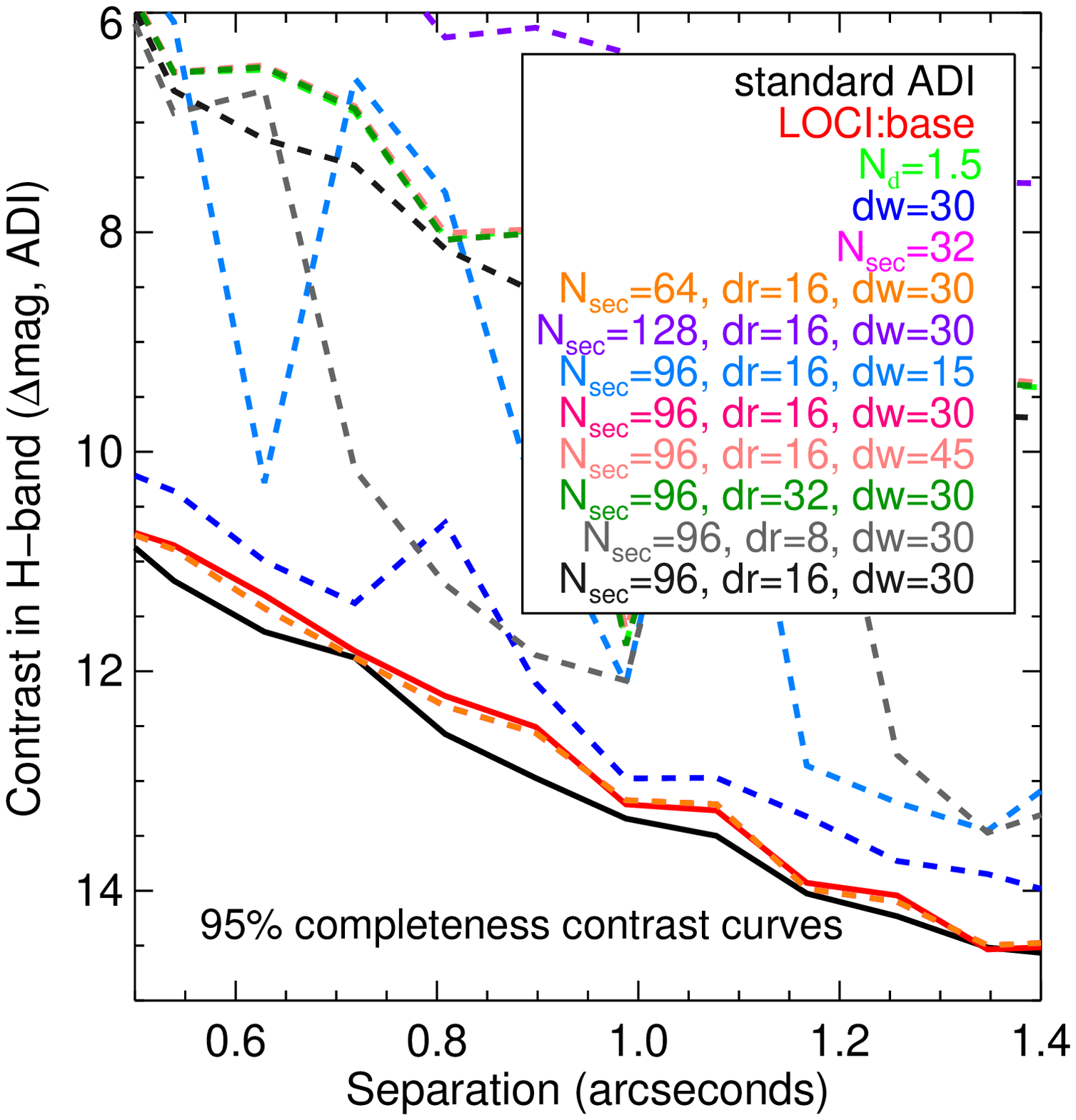}
      \includegraphics[height=8cm]{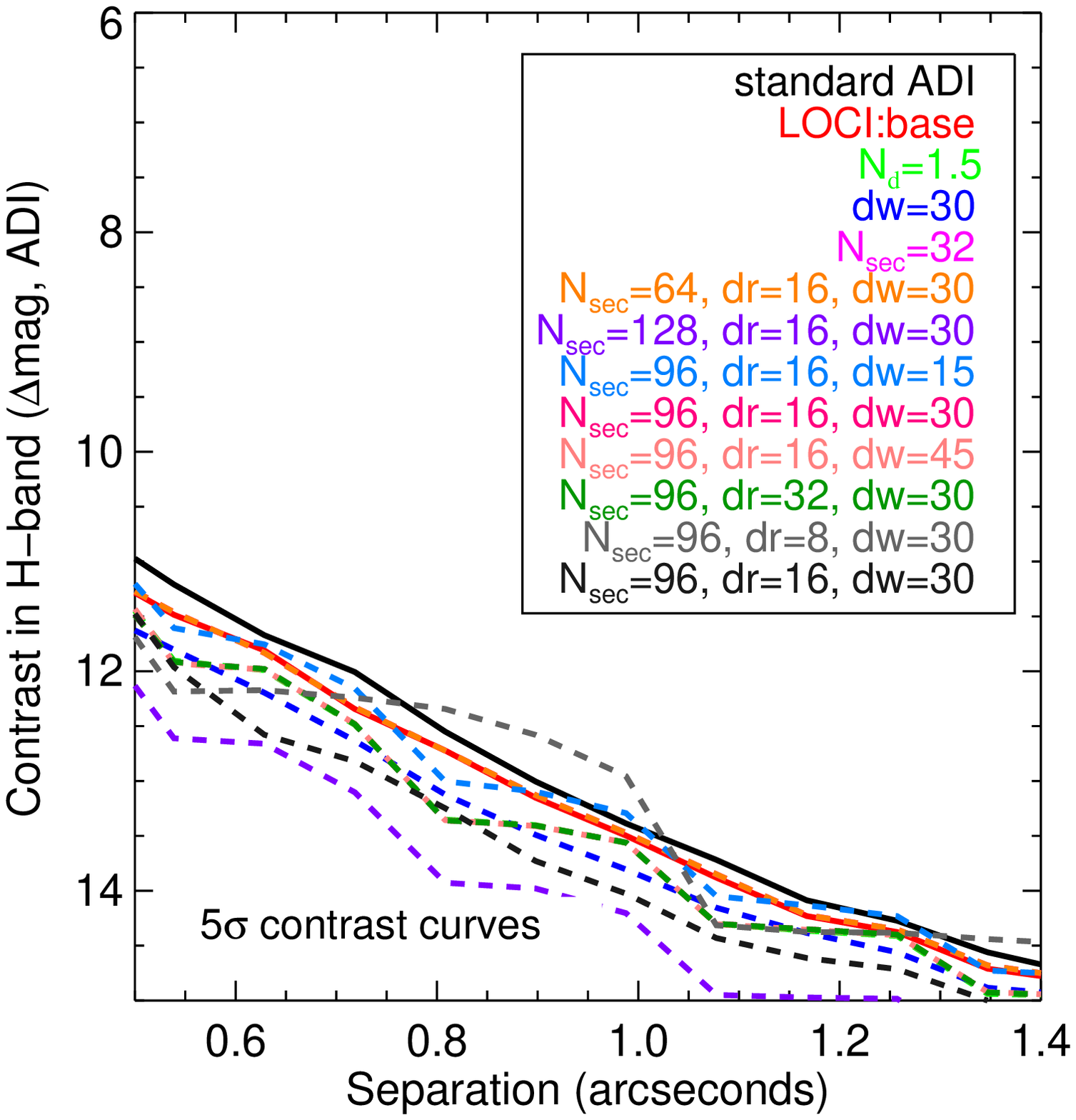}
    }
  \caption{Left: Comparison of 95\% completeness contrasts for different LOCI reductions and 
the standard reduction for UY Pic \textcolor{black}{one-channel} ADI 2009 data. The LOCI base parameters were set to 
$N_{sec}=8$, $N_{\delta}=3$,  $dr=4$,  $dr_{fac}=1.2$, $dw=1$ and filt = 'c' (see text for explanation).  
For the other LOCI reductions, we set some parameters to different values as noted in the legends. 
Also shown is the 95\% completeness contrast (black line) for the standard ADI algorithm. The LOCI reductions do not show  
any improvement over the standard ADI reduction. Right: Comparison of the nominal 5$\sigma$ contrast curves.}
\label{fig:loci_adi_95p_uypic}
\end{figure}

We tested the performance of several sets of parameters in the LOCI algorithm for our 
December 2009 UY Pic ASDI data set (45 images, each of 60 sec exposure time; total sky rotation of 34\dg). 
All combinations of parameters for $N_{sec} = \{8, 16, 32\}$, $dw=\{16,32\}$, $dr=\{2,4,6\}$ and  $N{_\delta}=\{0.5,1\}$ were evaluated.
However, no LOCI reduction yielded a noticeable improvement over the standard ASDI reduction, when comparing the 
95\% completeness curves. At the same time, we found that the LOCI reductions can produce final images with less noise.
In the left panel of Figure~\ref{fig:loci_asdi_95p_uypic}, we demonstrate the effect of changing each LOCI 
parameter from a base set of values. Although none of the LOCI 95\% completeness curves improve 
on the standard ASDI curve, in the right panel of Figure~\ref{fig:loci_asdi_95p_uypic} 
we see that the LOCI reductions yield better nominal 5$\sigma$ contrast curves. In the fact, the more 
aggressive LOCI reductions (smaller subtraction regions, etc.) yielded better nominal 5$\sigma$ contrast 
curves but worse 95\% completeness curves. Even for the plain
\textcolor{black}{one-channel} ADI UY Pic data set, the LOCI 95\% completeness curves 
showed no improvements over the standard reduction (Figure~\ref{fig:loci_adi_95p_uypic}). This result was confirmed 
when we took 45 ADI Campaign datasets from 2009 (see \S 7) and compared their 95\% completeness contrasts from 
LOCI and our standard ADI pipeline (Figure\ 11). 

Our results differ from those of \citet{2007ApJ...660..770L} perhaps
because we use a detection criteria \textcolor{black}{to define} our
completeness curve, \textcolor{black}{rather than just the radial noise
profile}. In their work, \citet{2007ApJ...660..770L} compare the
signal-to-noise of recovered simulated companions for LOCI and
non-LOCI algorithms but do not set a criteria for what counts as a
detection, or consider the false-positive threshold. Another 
possible explanation for the different results is that the higher
Strehl ratios achieved with NICI (Chun et al.\ 2008) have yielded
more stable quasi-static speckle patterns than obtained with the Gemini/ALTAIR AO system for GDPS. In this case, the special 
selection of reference frames with LOCI would not be very effective, because the reference frames are all comparably good 
matches. Experimentation with NICI data sets \textcolor{black}{longer than the Campaign
stardard of 45 mins}, where the quasi-static pattern 
changes significantly over the course of the observation, might show that in these cases, LOCI performs better. However, 
that experiment is beyond the scope of this paper. 

\citet{2007ApJ...660..770L} have already explained that LOCI can remove signal 
from real objects in the field if applied too aggressively. 
When the reference regions are too small or the reference images too numerous, the 
application of LOCI is too aggressive. In practice, even when reference images are bad 
matches to the science image, as the number of images is increased, LOCI keeps reducing 
the noise in the output image. Thus it is advisable to check that
simulated companions that are brighter than the contrast limits 
are not lost in the reduction process. Using the same method, one can also check that the contrast 
of the simulated companions are correctly recovered \citep{2010SPIE.7736E..52M}.

\section{ANALYSIS OF NICI CAMPAIGN DATA FROM 2009A}

\textcolor{black}{Since the NICI Campaign began in December 2008}, we have obtained 172 ASDI data sets and 241 ADI data sets as of August 2012. Since 
performing the simulated companion tests on all the Campaign data is
time-consuming, we limit our analysis to \textcolor{black}{45 ASDI and 45 ADI data sets} obtained in
the beginning of 2009. Some NICI targets have either only an ASDI or only an ADI observation, and so 
some of the targets in our ASDI and ADI sample data sets are different. We chose the standard ASDI reduction method over LOCI, since we noticed no 
improvement with the latter method (\S6). \textcolor{black}{
Figure~\ref{fig:2009_adi_95p_loci_comp} also shows that the NICI standard ADI pipeline and the LOCI pipeline gives the same performance for the 2009 ADI data sets.} 
As before, we generate the 95\% completeness contrasts for all our data sets.

\subsection{Achieved raw contrasts}
\begin{figure}[ht]
  \centerline{
      \includegraphics[height=8cm]{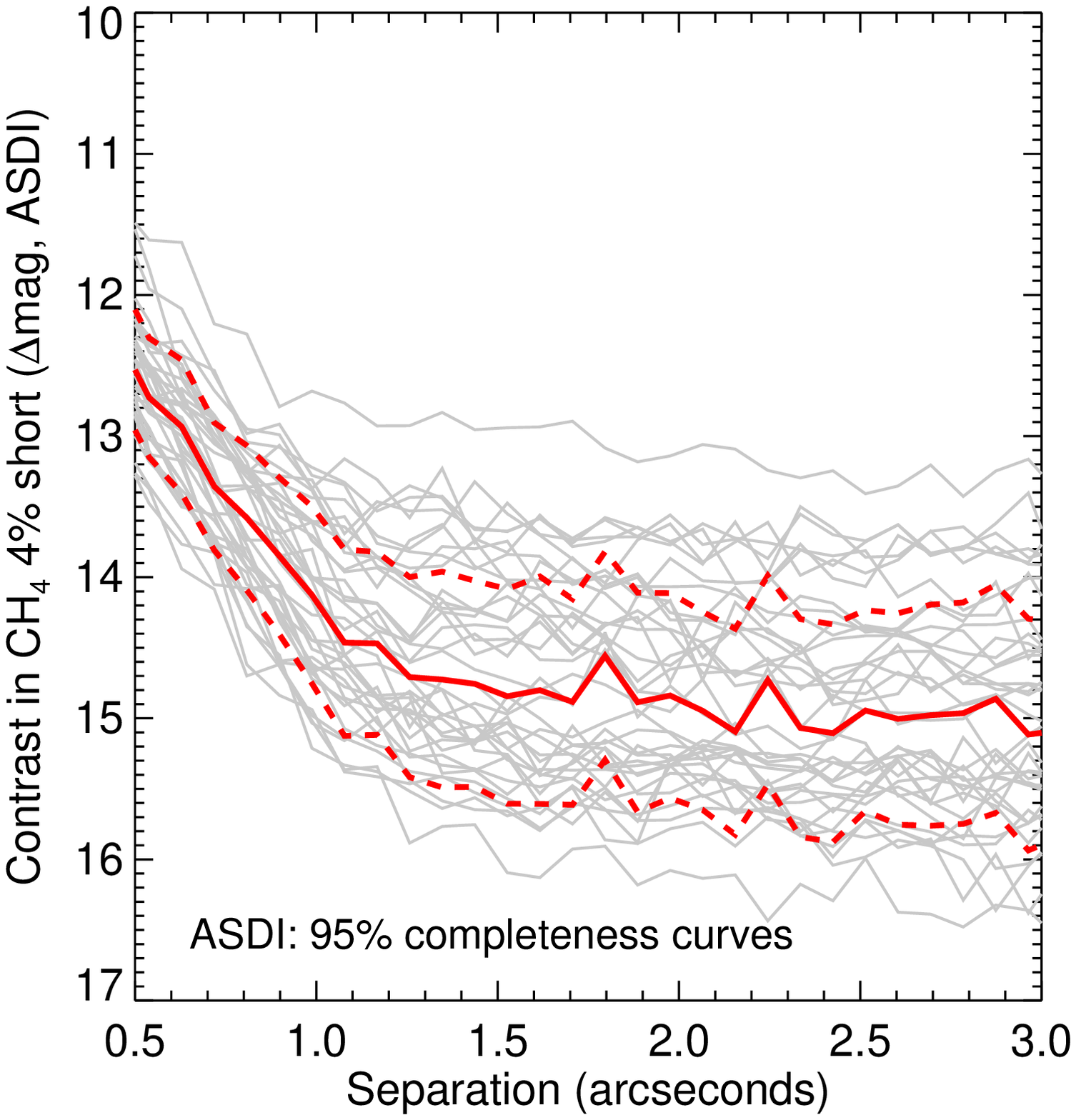}
      \includegraphics[height=8cm]{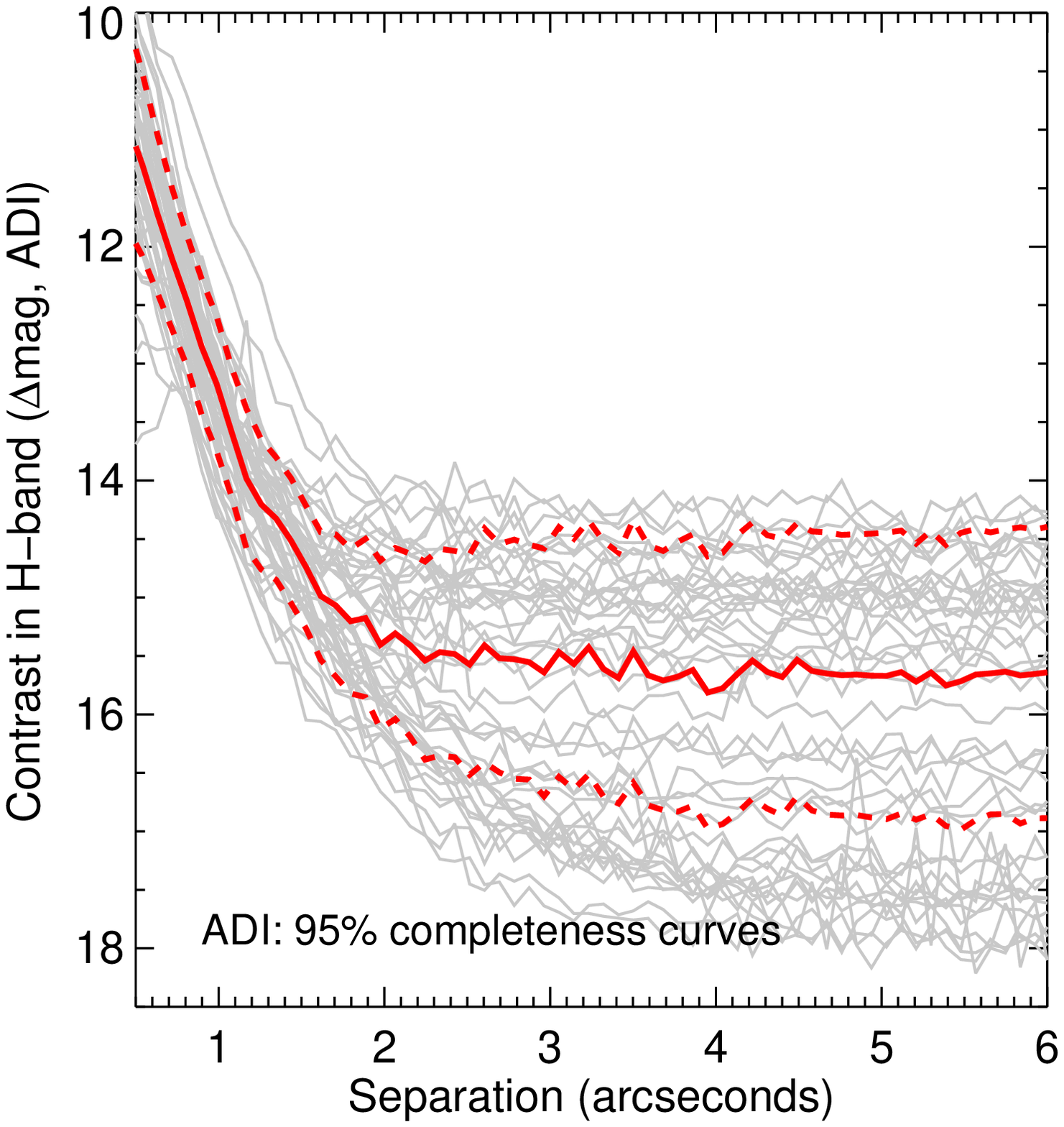}
    }
  \caption{Left: 95\% completeness contrasts from 45 NICI ADSI data sets obtained in 2009. Completeness 
contrasts from individual data sets are shown in gray. The median completeness curve is drawn as a solid red line, while the 1$\sigma$ range is shown by the dashed red lines. Right: 95\% completeness contrasts from 45 NICI ADI data sets obtained in 2009. Some of the stars in the ASDI and ADI samples are different.}
\label{fig:2009_95p}
\end{figure}

\begin{figure}[ht]
  \centerline{
      \includegraphics[height=8cm]{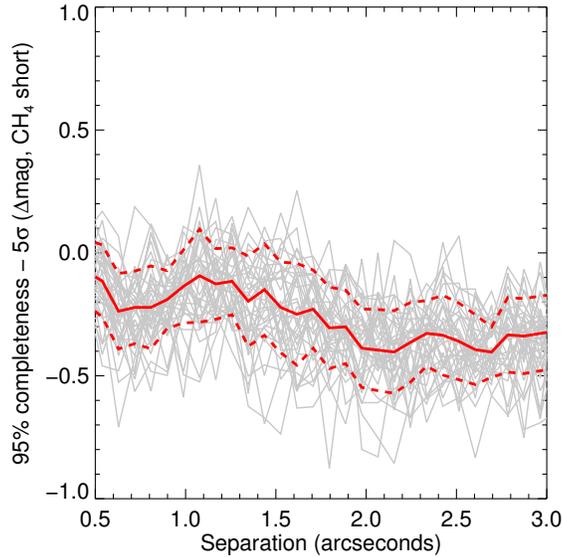}
    }
  \caption{The difference between the 95\% completeness and the 5$\sigma$ constrast curves for 45 NICI ADSI data sets obtained in 2009. The contrast 
differences from individual data sets are shown in gray. The median difference curve is drawn as a solid red line, while the 1$\sigma$ range is shown by the dashed red lines.}
\label{fig:2009_diff}
\end{figure}

\begin{figure}[ht]
  \centerline{
      \includegraphics[height=8cm]{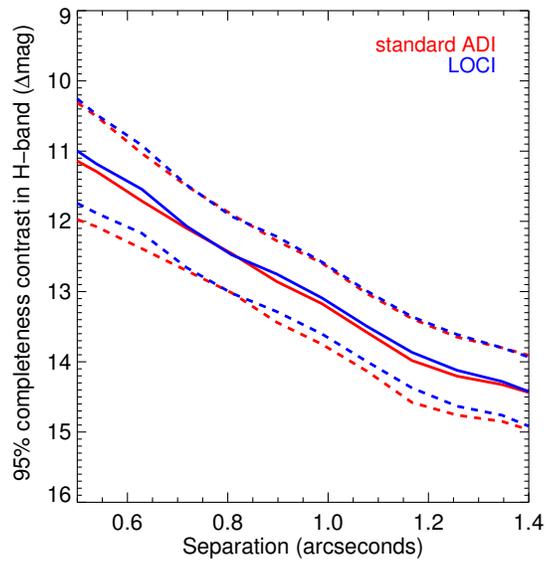}
    }
  \caption{Comparison of median contrast curves from the NICI standard ADI reduction (red) and the best LOCI reduction (blue) for 45 NICI ADSI data sets obtained in 2009.   The median completeness curves are drawn as solid lines, while the 1$\sigma$ range is shown by the dashed lines.}
\label{fig:2009_adi_95p_loci_comp}
\end{figure}

Figure~\ref{fig:2009_95p} shows all the \textcolor{black}{individual ASDI and ADI} 95\% completeness curves and also the median curve and the 1$\sigma$ dispersion.
\textcolor{black}{For this set of stars  ($V=$~4.3--13.9~mag)}, we should have recovered 95\% of
companions with a median ASDI contrast less than 12.6 mags
at 0.5$''$ and less than 14.4 mags at 1.0$''$.   The difference between the 95\% completeness 
curves and the nominal 5$\sigma$ curve for each dataset is displayed in Figure~\ref{fig:2009_diff} for comparison. The median 95\% curve is poorer than the median 5$\sigma$ 
curve by 0.1 and 0.3 mag at 0.5$''$ and 3$''$, respectively, varying slowly with separation. The difference is smallest at separations dominated by 
the stellar halo, increases with separation, and is roughly flat at separations beyond 2$''$.

At low temperatures ($<$1400~K), substellar objects with spectral type T \citep{2003A&A...402..701B}, 
exhibit methane absorption in their spectra, such 
that their red channel flux would be suppressed relative to their blue
channel \textcolor{black}{flux.
However,} these considerations only become important when we are trying to calculate companion mass sensitivities. 
So, we defer such a discussion to \S\ref{section:contrast_adjust_sdi}. Here, we just note that even for an 
object with no methane, there is little self-subtraction from the red channel beyond separations of 1$''$, since we radially stretch the 
red channel image before SDI subtraction. 
Inside of 1$''$, for objects without any methane,  the reduction of the ADI-mode  dataset or the separate ADI reductions of the red and the blue channels
should yield better sensitivities as the red channel flux is not subtracted in these.

Beyond a separation of 1.5$''$, the ADI $H$-band contrast curves are more sensitive, with 95\% completeness contrasts 
of 15.0 and 15.5 mag at 1.5$''$ and 3$''$, respectively. Beyond 3$''$, the ADI median completeness curve flattens out, as the noise from the sky background and detector readout begin
to dominate that from the halo. Note that at large separations the ADI contrast distribution in Figure\ 9 is bimodal. This is because for faint stars, we reach the background limit at smaller angular separations than for bright stars, and the brightness distribution of our primary stars is also roughly bimodal. The distribution of contrasts at large separations is also skewed such that the mean contrast (16.0 mag) is greater than the median contrast (15.5 mag). 
  
\subsection{Contrast curve adjustments}
\label{section:contrast_adjust}

The simulated companions used to estimate the contrast curves 
\textcolor{black}{are not subject to the} same alterations to their flux
that real objects in the field \textcolor{black}{are}. To adjust the contrast curves for
these alterations, we take into account three effects: (1) smearing 
of an object's flux as the field rotates during each individual exposure; (2) the non-zero mask opacity from 0$''$ to 
0.55$''$ separation from the primary; and (3) the flux loss from the red and blue channel differencing for objects that do not have photospheric methane absorption.

Note that another important effect, the contrast alteration due to ADI self-subtraction, is already accounted for in the 95\% completeness curves, 
since the simulated companions used to construct the curve undergo the same alteration  as any real object.
 
\subsubsection{Smearing correction} 
The representative area of a PSF core is $\pi a^2$, where $a$ is the half \textcolor{black}{width} at half \textcolor{black}{maximum} of the PSF, 
or about 1.5 NICI pixels. Due to sky rotation during an individual exposure, the flux in the PSF core is 
spread over an area $\pi a^2 + 2 a L$ \citep{2007ApJ...670.1367L}. Here, $L$ ($ = R \omega \Delta t$) 
is the arclength \textcolor{black}{traced by a real object during an individual exposure}, 
$\Delta t$ ( typically 1 minute for NICI observation). The separation from the primary and 
the rotation rate in radians per minute are $R$ and $\omega$, respectively.  Thus,  there is a fractional 
flux loss of \textcolor{black}{$1 + (2 R \omega \Delta t) / (\pi a)$} in the PSF core. We convert this loss into magnitudes as a 
function of separation and subtract it from the 95\% completeness contrasts. Out of 413 NICI observations obtained up to August 2012,  
382 (93\%) of NICI observations have an average rotation rate of less than 2 degrees per minute. 
Thus, 93\% of the time the contrast is degraded by less than 0.015 and 0.08 magnitudes at 1$''$ and 5$''$, respectively.

To check the reliability of the flux correction, we simulated the
smearing of Gaussian PSFs (FWHM= 3 pixels) for the
\textcolor{black}{target} declinations and observing hour angles
that yield very fast and non-uniform rotation rates. We tested declinations from -24\dg\ to -36\dg\ (Gemini South latitude$=-$30.24\dg)  
in steps of 2\dg\ and starting hour angles of $+$10 and $-$10 minutes.  
These choices yielded average rotation rates of 1 to 3.5\dg\ per minute. 
We placed the PSF at a separation of 3$''$ from the rotation center.
\textcolor{black}{For larger separations, one can use the superior ADI 
detection limits obtained with small rotation rates and thus avoid the smearing problem.}
We then measured the 3-pixel (radius) aperture flux of the median-combined 45 frames
of smeared PSFs from the varying sky rotations. This aperture flux was adjusted for the flux loss fraction we computed earlier 
and compared to the true aperture flux of the Gaussian PSF. We found in all cases that the fractional difference was less than 1.5\%. 
This precision is high enough for our purposes.

\subsubsection{Mask opacity correction} 
As described in \citet{2011ApJ...729..139W}, we observed a pair of stars, 
(2MASS~J06180157--1412573 and 2MASS~J06180179--1412599) separated by only a few arcseconds, to measure the opacity 
of the coronagraphic mask in the $CH_4$ 4\% short and long filters. The two stars were observed simultaneously 
without the mask and then observed with the brighter object at different displacements along a straight line from the center of the mask. 
We measured the contrast between the stellar pair
in the off-mask images and in the mask-occulted images, 
using an aperture radius of 3 pixels.
The difference between the on and off-mask contrasts (in magnitudes)
gives us the mask opacity \textcolor{black}{between} 
0$''$ to 0.55$''$ from the center of the mask. The measured opacities of the 0.32$''$ (radius) mask 
as a function of separation are given  in Table~\ref{tab:mask_opacity}.
At the center of the mask, the opacity is 6.39\pp0.03 mag, while at a separation of 0.55$''$ the mask becomes completely transparent.
\textcolor{black}{The uncertainty in the opacity is dominated by the
  uncertainty in the astrometry of a detected source through an
  opacity gradient. Our measurements have shown this astrometric
  uncertainty to be roughly 0.5 NICI pixels (9~mas). The opacity
  uncertainties given in Table~\ref{tab:mask_opacity} are the standard
  deviation of opacity values over 9~mas ranges.}

\textcolor{black}{
For the $H$-band, used in the ADI observations, we only have measurements of the central mask opacity. 
Thus for targets where we only conducted ADI observations, the $CH_4$ mask opacity adjustment is used, 
assuming that the $H$-band transmission profile has the same shape. 
The final sensitivity limits for a target star only depend on the better of the ASDI and ADI contrasts achieved at each separation.
At separations of less than 0.55$''$, the  ASDI observations taken
with $CH_4$ filters always yield superior contrasts, so our lack of
independent data for the $H$-band does not impact our survey results.}
%The beta pic measurements were done for the 0.22'' mask.

\subsubsection{SDI self-subtraction correction}
\label{section:contrast_adjust_sdi}

In the ASDI reduction, when the red channel image is demagnified to matched the speckles in the blue channel image,
at large enough separations ($>$1.2$''$) the red channel counterpart of a real object will almost completely shift off the blue channel counterpart.
However, at small separations the red and blue channel counterparts will still overlap.
Thus during image differencing, some flux from the blue channel will be removed and this has to be adjusted for.    
Of course, this SDI self-subtraction correction does not affect the ADI contrast curves.
We do not apply this correction to the ASDI contrast curves for the Campaign. 
Instead, we apply the correction when converting the ASDI contrast curves into 
companion-mass sensitivity curves, for which we have to assume a spectral type (e.g., T dwarf).
Monte Carlo simulations are used to determine the planet detection probability of each 
NICI observation \citep{2013ApJ...776....4N}. In these 
simulations thousands of companions of are drawn from a power-law distribution in mass.
Using evolutionary models \citep[e.g.][]{2003A&A...402..701B}, 
the temperatures corresponding to the masses of the simulated companions are computed, given the age of the host star.     
Next, we calculate the red to blue channel flux ratio given the
temperatures of the companions \textcolor{black}{and assumed spectral type to temperature relation}. 
Thus, we can estimate the ASDI contrasts of the companions and what
fraction of them are expected to be detected \textcolor{black}{given the measured contrast curve}. 
In other words, the SDI correction is not applied to the contrast curves themselves
but to the model-derived contrasts of simulated companions. 

The fractional flux loss due to SDI subtraction is a function of separation.
We estimate this loss by subtracting a Gaussian PSF of width 3 pixels from itself after shifting 
it radially by 4.6\% of the separation (which corresponds to red image demagnification) 
and taking the ratio of its fluxes before and after the subtraction. 
This maximum fractional loss is for objects without methane absorption. 
\textcolor{black}{For T dwarfs, the red channel has less flux than the blue channel due to methane 
absorption, and thus the loss is less severe for these objects.}

\textcolor{black}{
Using the spectrum of 2MASS~J04151954$-$093506 \citep[spectral type T8][]{2004AJ....127.3553K,2006ApJ...637.1067B},
we estimate that companions of spectral type T8 have a red to blue channel flux ratio of 0.125.
Thus, during the ASDI reduction process, at small separations roughly 0.125 mag of flux may be removed when subtracting the 
red channel from the blue channel.   
To recover the true CH$_4$ 4\% short (blue channel) sensitivity to a T8 at small separations, we 
need to decrease the contrast estimate from the ASDI reduction by 0.125 mags.}
 
We find that even with significant self-subtraction, the ASDI reduction can yield superior contrasts to the ADI reductions.
For example, for a companion with no methane, the ASDI contrasts at 
0.5$''$, 0.75$''$ and 1.0$''$ separation are reduced by 0.58, 0.49 and 0.2 mags, respectively, relative to its blue channel contrasts.
Comparing the median contrast curves in Figure~\ref{fig:2009_95p}, we note that the ASDI curve is more sensitive 
than the ADI curve by 1.5, 1.35 and 0.9 mags at the same separations, respectively.
Thus at small separations, the ASDI contrast curves are \textcolor{black}{typically more} sensitive, whether the companions are methane-bearing or not.

\subsection{Photometric accuracy}
\label{section:photast}
\textcolor{black}{Our image reduction procedure introduces small systematic and random errors into the 
photometry of detected objects.} Here, we estimate there errors by studying the recovered 
simulated companions. The original simulated companions were inserted into the individual frames as 20$\sigma$ 
sources relative to the $1\sigma$ contrast curve (\S~4.3). 
When these are recovered after the reduction procedure, their
\textcolor{black}{measured} contrast with respect to the 
primary is \textcolor{black}{differs from the original value}. 
The differences between the input contrasts and the recovered contrasts for the 2009 ASDI data sets are plotted 
against separation in Figure~\ref{fig:phot_ast_acc}. At each separation, we plot the median 
of the contrast \textcolor{black}{difference} as a solid line and indicate the standard deviation by two dashed lines. 
The contrast \textcolor{black}{difference} increases from 0.05 to 0.3 mags between 0.5 and 2$''$ and settles to 0.25 mags 
at 6$''$. The standard deviation suggests a uniform photometric uncertainty of 0.07 mags between 0.5 and 6$''$.
\textcolor{black}{We use these results to correct the photometry of detected companions. 
(Note that the 95\% completeness curves do not require this correction.)} 

\subsection{Astrometric accuracy}
The reduction procedure also introduces some astrometric uncertainty in the location of companions 
with respect to the primary. \textcolor{black}{We plot the astrometric displacement between the original 
locations of the artificial companions and their recovered locations in Figure~\ref{fig:phot_ast_acc}.
(The uncertainty is of course expected to be smaller for brighter companions.)}
We see that the astrometric uncertainty is roughly uniform 
($\approx$~10~mas) as a function of separation. However, we must also include the uncertainties 
introduced by centroiding, residual distortions and the plate scale estimate. 

The astrometric uncertainty due to the precision limit of our
centroiding algorithm, given the finite size of the NICI pixels, is 0.011 pixels (0.2 mas).
We estimate this by testing our centroiding algorithm on Gaussian
point sources \textcolor{black}{placed in a noise-free image} with different sub-pixel offsets relative to the pixel center.   
The standard deviation of the difference between the known positions
and the measured positions of the point sources gives the uncertainty
of the centroiding algorithm.

For saturated datasets, we compare the centroids of the short exposure unsaturated images, taken just prior to the science data, to the centroid estimate 
of the first saturated image. The movement of the star relative to the mask should be very small for the minute of elapsed time between the exposures.
From analyzing several saturated datasets, we estimate that the
centroiding uncertainty for saturated images is 0.5 pixels or 9~mas.
   
\begin{figure}[ht]
  \centerline{
    \hbox {
      \includegraphics[height=8cm]{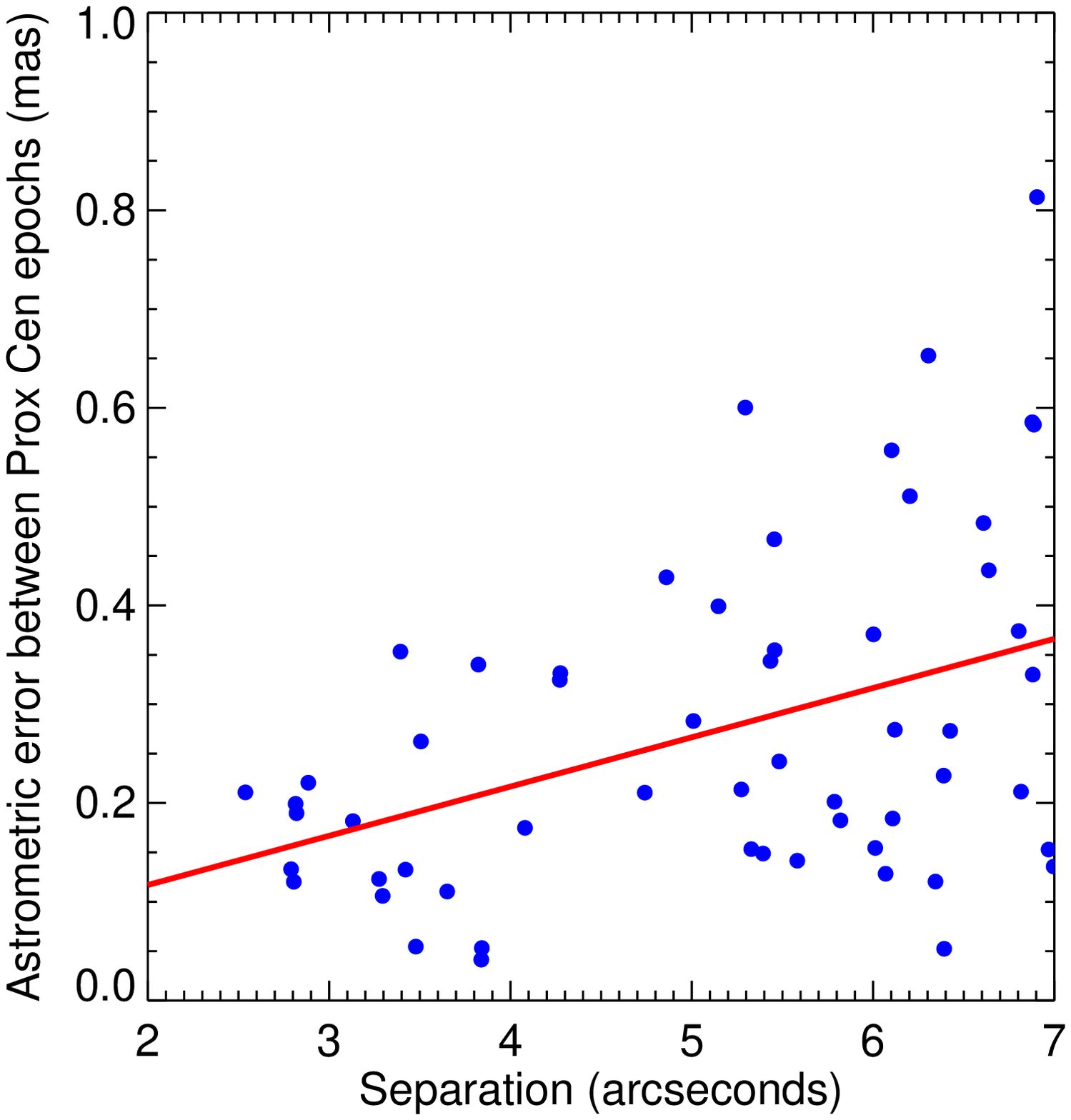}
      \includegraphics[height=8cm]{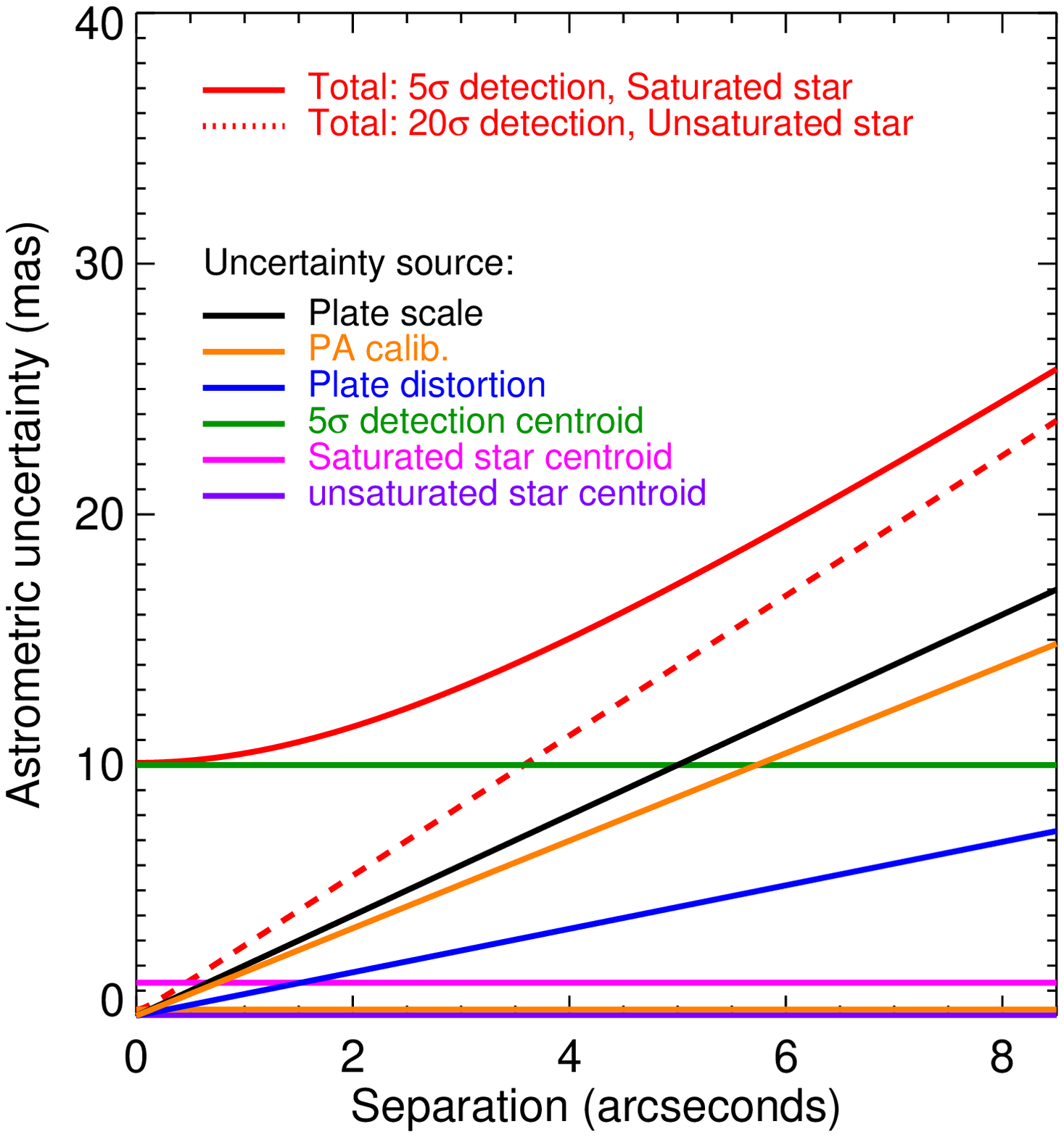}
    }
  }
  \caption{Left:  Astrometric error between two epochs of Prox Cen NICI observations taken on UT 2009 April 8 and April 26. The 57 detections with signal-to-noise above 
30 are shown. These were used to produce a robust line-fit for error 
vs separation (red line). Right: NICI astrometric errors from all the 
known sources as a function of separation (see text). The total errors 
for two cases are shown: (1) sources found at the 
detection limit in a dataset with a saturated primary, and (2) sources detected at the 20$\sigma$ level around an unsaturated star.}
  
  \label{fig:astrom_parts}
\end{figure}

\begin{figure}[ht]
  \centerline{
    \hbox {
      \includegraphics[height=8cm]{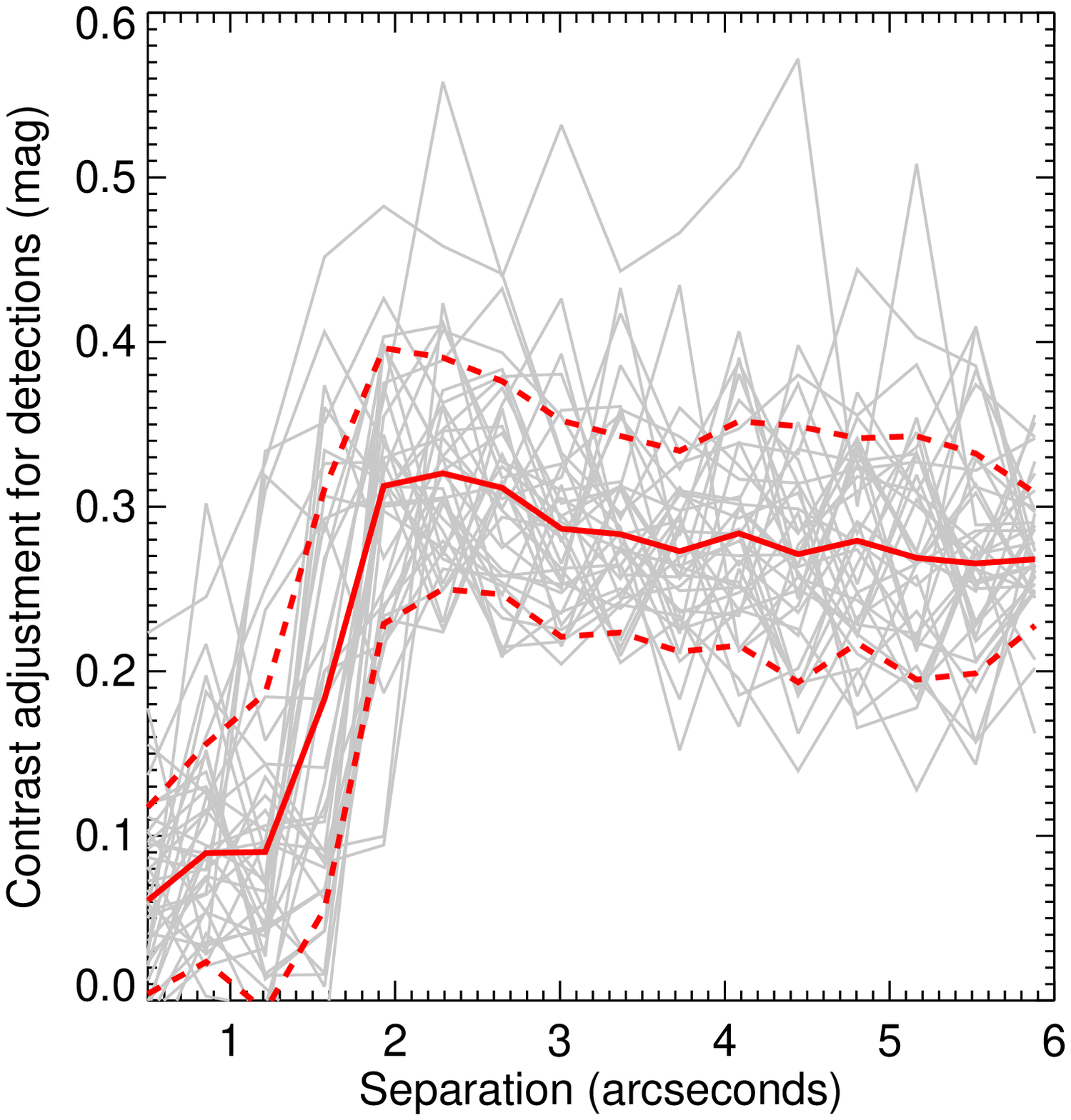}
      \includegraphics[height=8cm]{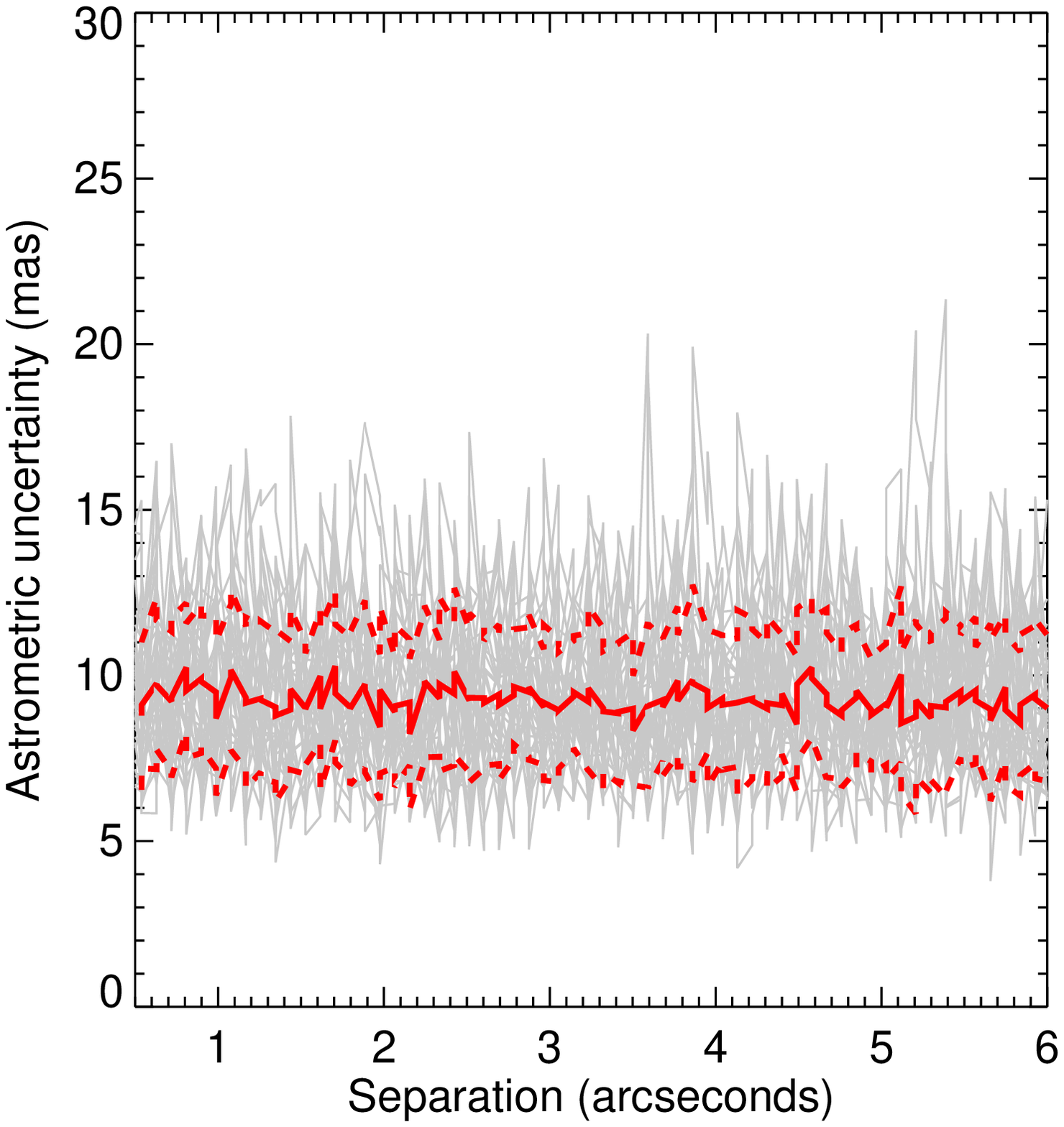}
    }
  }
  \caption{Left: Contrast adjustment to be applied to the photometry
    of detected companions. This adjustment or systematic error was
    estimated from 45 ASDI data sets from 2009. The adjustments for
    individual data sets are shown in gray. The median adjustment and
    the standard deviation are shown as a function of separation in
    red solid and dashed lines, respectively. Right: The astrometric
    uncertainties estimated from the same reductions. Individual,
    median and standard deviation curves are shown in the same color
    scheme as the \textcolor{black}{left figure}. }
  
  \label{fig:phot_ast_acc}
\end{figure}

The astrometric uncertainty due to imperfect distortion correction was found to be 0.017+0.049$\rho$ pixels, where $\rho$ is the projected separation in arcseconds.
This was measured by comparing the detections in the Proxima Centauri field from two NICI epochs obtained on 2009 April 8 and April 26 UT. 
Out of  73 background sources, the differences in the positions of 57 sources with S/N$>$30 were 
fitted robustly with a line (Figure~\ref{fig:astrom_parts}), which
resulted in the solution above.
  
The red and blue channel plate scales were measured from NICI
observations \textcolor{black}{of a field in} the Large Megallanic Cloud. This field has been observed with Hubble Space Telescope Advanced Camera for Surveys 
to an astrometric precision of 1 mas (Diaz-Miller 2007).
The astrometric uncertainty in the \textcolor{black}{NICI} plate scale measurement is estimated to be 0.2\%.
From these observations, we also estimated that the uncertainty in the direction of North is 0.1 degrees. 

The total astrometric error is all these contributions added in quadrature.  Figure~\ref{fig:astrom_parts} shows all
the contributions as a function of separation. The total astrometric uncertainty for a 20$\sigma$ source detection around 
an unsaturated star is 2.2 and 11 mas at 1$''$ and 5$''$, respectively. The total astrometric uncertainty for a 5$\sigma$ source detection around 
a saturated star is 10 and 15 mas at 1$''$ and 5$''$, respectively.

\section{CONCLUSIONS}
We have presented the image processing techniques used to detect companions to target stars in the Gemini NICI Planet-Finding 
Campaign. \textcolor{black}{We have achieved} median contrasts of 12.6 magnitudes at 0.5$''$ and 14.4 
magnitudes at 1$''$ separation, for a sample of 45 stars
\textcolor{black}{($V=$~4.3--13.9 mag)} from the early part of our campaign. 
We have also \textcolor{black}{implemented a novel technique to
calculate completeness contrast curves, accounting 
for both completeness and false positives.}
We have demonstrated how this technique can be used to determine which 
image-processing pipelines are more successful at recovering faint simulated point sources. Specifically, 
we analyzed the effect of different image filters on the final contrast curves. We have compared our pipeline 
to the performance of the LOCI algorithm on NICI data, and found that LOCI is not 
more effective than our standard pipeline. 

Our processing technique is conceptually similar to previous techniques, \textcolor{black}{but differs in that} we use special image filters and 
fit speckle patterns prior to reference PSF subtraction. The next-generation exoplanet finders, 
SPHERE \citep{2010ASPC..430..231B} and the Gemini Planet Imager \citep{2008SPIE.7015E..31M}, 
will both be equipped with integral field spectrographs \textcolor{black}{(also see
\citet{2013ApJ...768...24O}; Project 1640 and
\citet{2011SPIE.8149E...7G}; SCExAO)} and will thus require more complex
algorithms to isolate the exoplanet signals \citep[e.g.][]{2012ApJS..199....6P}. However, for the faintest planets, some version of the 
ASDI method discussed here will probably have to be used as the signal will be too weak to obtain resolved spectra. 
Thus, the techniques demonstrated here may be instrumental in addressing exoplanet detection challenges of the future.    

\acknowledgements

This work was supported in part by NSF grants AST-0713881 and AST- 0709484
awarded to M. Liu and NASA Origins grant NNX11 AC31G awarded to
M. Liu.  The Gemini Observatory is operated by the Association of Universities for Research in
Astronomy, Inc., under a cooperative agreement with the NSF
on behalf of the Gemini partnership:
the National Science Foundation (United States), the Science and Technology Facilities Council
(United Kingdom), the National Research Council (Canada), CONICYT (Chile), the Australian
Research Council (Australia), CNPq (Brazil), and CONICET (Argentina). 
Our research has employed the 2MASS data products; NASA's Astrophysical
Data System; the SIMBAD database operated at CDS, Strasbourg, France.
%;
%the M, L, and T~dwarf compendium housed at DwarfArchives.org and
%maintained by Chris Gelino, Davy Kirkpatrick, and Adam Burgasser; and
%the SpeX Prism Spectral Libraries maintained by Adam Burgasser at
%http://www.browndwarfs.org/spexprism. 

{\it Facilities:} Gemini-South (NICI).

\vfill
\eject

\bibliographystyle{apj}
\bibliography{zrefs}

\vfill
\eject

\begin{deluxetable}{ccc}
\tablecaption{Relation between LOCI parameters}
\tablewidth{0pt}
\tablehead{ \colhead{$N_{sec}$, $dr_{fac}$, $dw$} &\colhead{$N_A$, $g$ at $r=0.5''$} & \colhead{$N_A$, $g$ at $r=7''$} }
\startdata
8, 1.2, 16  & 690, 0.1 & 3400, 0.3 \\
8, 1.4, 16 &  570, 0.1 & 4200, 0.35 \\
32, 1.2, 16  & 340, 0.2 & 960, 1.1 \\
32, 1.4, 16  & 320, 0.2 & 1000, 1.5 \\
8, 1.2, 1  & 160, 0.04 & 2700, 0.2 \\
8, 1.4, 1 &  160, 0.04 & 3500, 0.3 \\
32, 1.2, 1  & 80, 0.07 & 690, 1.0 \\
32, 1.4, 1  & 70, 0.07 & 840, 1.3 
\enddata
\label{tab:loci_prams}
\end{deluxetable}

\begin{deluxetable}{ccc}
\tablecaption{Measured opacities of the 0.32$''$ (radius) mask as a function of separation.}
\tablewidth{0pt}
\tablehead{\colhead{Separation ($''$)} & \colhead{$\Delta CH_4$
    short (mag)} & \colhead{ uncertainty (mag)}}
\startdata
0.0 & 6.39 & 0.03\\
0.2 & 4.5   & 0.22\\
0.3 & 2.1   & 0.09\\
0.4 & 0.73 & 0.02\\
0.55 & 0.0 & - \\
\enddata
\label{tab:mask_opacity}
\end{deluxetable}

%%%%%%%%%%%%%%%%%%%%%%%%%%%%%%   FIGURES  %%%%%%%%%%%%%%%%%%%%%%%%%%%%%%%

%FIGURE: LOCI SECTORS

% FIGURE: LOCI ASDI 95% contrasts UY Pic

% FIGURE: LOCI ADI 95% contrasts UY Pic

% FIGURE: 2009 ASDI and ADI 95% contrasts 

% FIGURE: 2009 ASDI. Difference between the 95% and 5 sigma curves 

% FIGURE: 2009 ADI  LOCI and NICI standard 95% contrasts comparisons for median and 1 sig variance curves 

\end{document}